\pdfoutput=1
\documentclass[%
    sigconf,
    nonacm
]{acmart}

\usepackage[english]{babel}
\usepackage{blindtext}
\usepackage{comment}
\usepackage{url}
\usepackage{multirow} %
\usepackage{xspace}
\usepackage{color, colortbl} 
\usepackage{cleveref} %
\usepackage{graphicx}
\usepackage{pifont} %
\usepackage{tikz}
\usepackage{bm} %
\usepackage{todonotes}
\usepackage{caption}
\usepackage{soul} %
\usepackage{ifthen}
\usepackage{algorithm}
\usepackage[noend]{algpseudocode} %
\usepackage{subcaption} %
\usepackage[nolist]{acronym}
\usepackage{listings}
\usepackage{csquotes}
\usepackage[inline]{enumitem}
\usepackage{mathtools}
\usepackage{nth}
\usepackage{circledsteps}
\usepackage{balance} %

\newboolean{showcomments}
\setboolean{showcomments}{false}

\newcommand\important[1]{\todo[inline]{\textbf{Important:} #1}}
\newcommand\gareth[1]{\textbf{\textcolor{blue}{GT: #1}}}
\newcommand\leo[1]{\todo[color=orange,inline]{\textbf{Leo:} #1}}
\newcommand\michal[1]{\textbf{\textcolor{magenta}{Michal: #1}}}
\newcommand\onur[1]{\todo[color=red,inline]{\textbf{Onur:} #1}}
\newcommand\maciej[1]{\todo[color=violet,inline]{\textbf{Maciej:} #1}}
\newcommand\saidu[1]{\todo[color=yellow,inline]{\textbf{Saidu:} #1}}
\newcommand\ignacio[1]{\todo[color=cyan,inline]{\textbf{Ign:} #1}}

\ifthenelse{\boolean{showcomments}} { }
{
\renewcommand\important[1]{}
\renewcommand\leo[1]{}
\renewcommand\michal[1]{}
\renewcommand\onur[1]{}
\renewcommand\maciej[1]{}
\renewcommand\saidu[1]{}
\renewcommand\ignacio[1]{}
\renewcommand\gareth[1]{}
}

\ifthenelse{\boolean{showcomments}}
{ \newcommand{\mynote}[3]{
    \protect\fbox{\bfseries\sffamily\scriptsize#1}
    {\small\textsf{\emph{\color{#3}{#2}}}}}}
{ \newcommand{\mynote}[3]{}}

\newcommand{\lb}[1]{\mynote{Leo}{#1}{blue}}

\newcommand{\cf}{cf.\@\xspace}

\newcommand{\etc}{etc.\@\xspace}

\newcommand{\etal}{\textit{et al.}\@\xspace}
\newcommand{\eg}{\textit{e.g.}\@\xspace}

\newcommand{\ie}{\textit{i.e.}\@\xspace}

\newcommand\para[1]{\vspace{0.05in} \noindent \textbf{#1.}}

\definecolor{verylightgray}{gray}{0.8}

\newcolumntype{L}{l<{\hspace{1cm}}}
\newcolumntype{C}{c<{\hspace{1cm}}}
\newcolumntype{D}{c<{\hspace{0.3cm}}}

\crefname{section}{Sec.}{Sec.}
\Crefname{section}{Section}{Sections}
\crefname{equation}{eq.}{eq.}
\crefname{figure}{Fig.}{Fig.s}
\Crefname{figure}{Figure}{Figures}

\floatstyle{plain}
\newfloat{lstfloat}{htbp}{lop}
\floatname{lstfloat}{Listing}
\crefalias{lstfloat}{listing}

\newcommand{\bsky}{Bluesky\xspace}
\newcommand{\bskyCo}{Bluesky PBC\xspace}

\usetikzlibrary{positioning}
\usetikzlibrary{graphs}
\usetikzlibrary{shapes.multipart}
\usetikzlibrary{shapes.misc}
\usetikzlibrary{matrix}
\usetikzlibrary{shapes}
\usetikzlibrary{arrows}
\usetikzlibrary{fit}
\usetikzlibrary{calc}
\usetikzlibrary{decorations.pathreplacing}

\tikzstyle{startstop} = [rectangle, rounded corners, minimum width=3cm, minimum height=1cm,text centered, draw=black, fill=red!30]
\tikzstyle{io} = [trapezium, trapezium left angle=70, trapezium right angle=110, minimum width=2.5cm, text width=2.5cm, minimum height=1cm, text centered, draw=black, fill=blue!30]
\tikzstyle{process} = [rectangle, minimum width=2.5cm, minimum height=1cm, text width=2.5cm, text centered, draw=black, fill=orange!30]
\tikzstyle{decision} = [diamond, aspect=3, minimum width=3cm, minimum height=1cm, text centered, draw=black, fill=green!30]
\tikzstyle{arrow} = [thick,->,>=stealth]
\tikzstyle{darrow} = [thick,<->,>=stealth]
\tikzstyle{entity_own} = [rectangle, minimum width=2.5cm, minimum height=1cm, text width=2.5cm, text centered, draw=black, fill=orange!30]
\tikzstyle{entity_foreign} = [rectangle, minimum width=2.5cm, minimum height=1cm, text width=2.5cm, text centered, draw=black, fill=gray!10]
\tikzstyle{functionality} = [rectangle, minimum width=2.5cm, minimum height=1cm, text width=2.5cm, text centered, draw=black, fill=white]
\tikzstyle{dataset} = [rectangle, minimum width=2.5cm, minimum height=1cm, text width=2.5cm, text centered, draw=black, fill=blue!10]
\tikzstyle{circlenode} = [circle, inner sep=0.05cm, draw=black, fill=gray!10, text centered]
\tikzstyle{ordernode} = [circle, thick, inner sep=0.02cm, draw=black, text centered]

\copyrightyear{2024}
\acmYear{2024}
\setcopyright{rightsretained}
\acmConference[IMC '24]{Proceedings of the 2024 ACM Internet Measurement Conference}{November 4--6, 2024}{Madrid, Spain}
\acmBooktitle{Proceedings of the 2024 ACM Internet Measurement Conference (IMC '24), November 4--6, 2024, Madrid, Spain}
\acmDOI{10.1145/3646547.3688407}
\acmISBN{979-8-4007-0592-2/24/11}

\makeatletter
\gdef\@copyrightpermission{
	\begin{minipage}{0.3\columnwidth}
		\href{https://creativecommons.org/licenses/by/4.0/}{\includegraphics[width=0.90\textwidth]{4ACM-CC-by-88x31.eps}}
	\end{minipage}\hfill
	\begin{minipage}{0.7\columnwidth}
		\href{https://creativecommons.org/licenses/by/4.0/}{This work is licensed under a Creative Commons Attribution International 4.0 License.}
	\end{minipage}
	\vspace{5pt}
}
\makeatother

\begin{document}

\begin{acronym}[Derp]
    \acro{did}[DID]{Decentralized Identifier}
    \acro{nsfw}[NSFW]{Not Safe For Work}
    \acro{atp}[ATProto]{Authenticated Transfer Protocol}
    \acro{pds}[PDS]{Personal Data Server}
    \acroplural{pds}[PDSes]{Personal Data Servers}
    \acro{uri}[URI]{Uniform Resource Identifier}
    \acro{diddoc}[DID Document]{DID Document}
    \acro{fqdn}[FQDN]{Fully-Qualified Domain Name}
    \acro{dns}[DNS]{Domain Name System}
    \acro{cbor}[CBOR]{Concise Binary Object Representation}
\end{acronym}

\title{Looking AT the Blue Skies of Bluesky}

\author{Leonhard Balduf}
\orcid{0000-0002-3519-7160}
\affiliation{%
	\institution{Technical University of Darmstadt}
	\city{Darmstadt}
	\country{Germany}}
\email{leonhard.balduf@tu-darmstadt.de}

\author{Saidu Sokoto}
\orcid{0000-0001-7152-546X}
\affiliation{%
	\institution{City, University of London}
	\city{London}
	\country{United Kingdom}}
\email{saidu.sokoto@city.ac.uk}

\author{Onur Ascigil}
\orcid{0000-0002-3023-6431}
\affiliation{%
	\institution{Lancaster University}
	\city{Lancaster}
	\country{United Kingdom}}
\email{o.ascigil@lancaster.ac.uk}

\author{Gareth Tyson}
\orcid{0000-0003-3010-791X}
\affiliation{%
	\institution{Hong Kong University of Science and Technology (GZ)}
	\city{Guangzhou}
	\country{China}}
\email{gtyson@ust.hk}

\author{Björn Scheuermann}
\orcid{0000-0002-1133-1775}
\affiliation{%
	\institution{Technical University of Darmstadt}
	\city{Darmstadt}
	\country{Germany}}
\email{scheuermann@kom.tu-darmstadt.de}

\author{Maciej Korczyński}
\orcid{0000-0002-4334-3260}
\affiliation{%
	\institution{University of Grenoble Alps} %
\city{Grenoble}
\country{France}
}
\email{maciej.korczynski@univ-grenoble-alpes.fr}

\author{Ignacio Castro}
\orcid{0000-0002-7739-6184}
\affiliation{%
	\institution{Queen Mary, University of London}
	\city{London}
	\country{United Kingdom}}
\email{i.castro@qmul.ac.uk}

\author{Michał Król}
\orcid{0000-0002-3437-8621}
\affiliation{%
	\institution{City, University of London}
	\city{London}
	\country{United Kingdom}}
\email{michal.krol@city.ac.uk}

\renewcommand{\shortauthors}{Leonhard Balduf et al.}


\begin{abstract}
The pitfalls of centralized social networks, such as Facebook and Twitter/X, have led to concerns about control, transparency, and accountability.
Decentralized social networks have emerged as a result with the goal of empowering users.
These decentralized approaches come with their own trade-offs, and therefore multiple architectures exist.
In this paper, we conduct the first large-scale analysis of \bsky, a prominent decentralized microblogging platform.
In contrast to alternative approaches (\eg Mastodon), \bsky decomposes and opens the key functions of the platform into sub-components that can be provided by third party stakeholders.
We collect a comprehensive dataset covering all the key elements of \bsky, study 
user activity and assess the diversity of providers for each sub-components.
\end{abstract}

\maketitle
\acresetall

\section{Introduction}\label{sec:intro}

Social media platforms like Facebook and Twitter have become ubiquitous, attracting vast user bases and wielding significant influence~\cite{raman2019challenges}. However, their centralized structure raises concerns about control, transparency, and accountability.
These concerns stem from the dominance of these platforms and their unchecked discretion over user behavior and content moderation, which has sparked regulatory interest and public debate~\cite{frier2021why, satariano2021facebook}.

In response, decentralized social network (DSN) platforms have emerged.
Aiming to foster a more democratic and open online environment, DSNs have pioneered diverse architectures with varying degrees of decentralization. 
The \enquote{fediverse}, with its server-based federated services like Mastodon, has attracted attention and users, particularly after the change of ownership in Twitter~\cite{he2023flocking}.
These platforms deconstruct their service into independent, user-creatable server instances. 
This approach to decentralization shifts control to instance administrators, who moderate content and users~\cite{rozenshtein2023moderating}. The downside of this is that the failure of an individual server can result in data and account loss~\cite{raman2019challenges}. 
Other approaches, such as Nostr~\cite{nostr}, overcome these problems by greater decentralized replication, but consequently lack important server-centric features such as content recommendations.

Bluesky~\cite{kleppmann2024bluesky} attempts to overcome these deficiencies. 
Deployed in 2022, Bluesky operates as a microblogging service that resembles Twitter/X where users can follow each other and share short posts (including images and videos). 
Bluesky, however, proposes a radical departure from Twitter/X or existing fediverse implementations.
Its key innovation is to decompose and open the key functions of a social microblogging platform into sub-components that can be provided by stakeholders other than \bsky. 
In contrast to fediverse applications, which embed all the functions into a single server, Bluesky encourages multiple stakeholders to take on the responsibility of delivering particular sub-aspects of the social media experience. This means that multiple actors can develop particular components within the overall system. Subsequently, users can then select between these competing providers of each component to compose their own personalized social media experience. 

To attain this, \bsky defines five key system components:
\begin{enumerate*}
\item Decentralized User Identifiers (DIDs): To detach users from any specific operator, each user can create their own distinct (cryptographically verified) identifier, which they can use across different providers. This identifier is linked to their user handle, a human-readable identifier with a domain of the user's choice.%
\item Personal Data Servers (PDSes): To detach users from relying on a specific server to host their data, users can port their data (associated with their DID) to any data server and even create their own PDS.
\item Relays: To minimize complexity and overheads, data from multiple PDSes can be aggregated within a single Relay server that offers high-performance delivery to end users. This is optional and clients can still retrieve posts directly from the PDSes.
\item Feed Generators: To allow a plurality of feed algorithms, anybody can develop their own Feed Generators, which define the selection and order of posts seen on a user's timeline. Importantly, users can select between competing Feed Generators to configure how they view content.
\item Labelers: To facilitate content moderation, anybody can develop a Labeler which assigns labels (\eg hate speech) to objects, including posts and accounts. These can also be used locally by clients to decide content that should be filtered.
\end{enumerate*}

Any competing operator can build the above components, and users can freely select between them. 
For example, users can migrate their data between competing PDSes or configure the use of alternative Feed Generators. This plurality constitutes a radical shift from the walled-garden approach espoused by existing major players and proposes novel innovations with respect to other DSNs.
\bsky produces a complex system of interconnected components, operators, and users with a diversity of experiences.
In this paper, we examine the operation of these components, 
study whether operators uptake the opportunity of providing them, whether this results in competing offerings and how users choose among offerings and operators.
We particularly focus on how multiple operators manage the critical components: PDSes, Feed Generators, and Labelers. With this in mind, we gather the first large-scale \bsky dataset, covering $5{,}523{,}919$ users, $225{,}461{,}969$ posts, $40{,}398$ Feed Generators, and $62$ Labelers.

We start by investigating the competing set of domains hosting handles.
We discover that, despite the supposed openness, the vast majority of users' handles are linked to bsky.social. 
We then inspect the availability and offerings of Labelers.
Although \bsky operates the most popular, we do observe a growing ecosystem of operators now issuing a majority of labels just two months after opening this system component.
We find that some community operators deviate from the original Labeler goal of filtering content and rather focus on content tagging.
We then investigate the Feed Generators. This is the most popular and competitive component of the ecosystem, with tens of thousands of active Feed Generators in operation.
We discover a wide range of feeds, from spam accounts to ones dedicated to explicit content. To the best of our knowledge, this is the first large-scale study of the \bsky component ecosystem.

\section{Bluesky Primer}\label{sec:background}
\bsky is a decentralized network where each system component (\eg data storage, moderation engine) is open-source and can be replicated, operated, and modified by the community.
Users have control over their data, and can migrate between competing operators freely.
\bsky attempts to be scalable and easy to use, by avoiding redundant communication and providing a user experience similar to centralized social networks.
We use \bsky to refer to the social network platform and \bskyCo when referring to the company developing the platform.
In this section, we provide an overview of the key concepts of Bluesky.

\para{The AT Protocol}
\label{sec:background:components:atproto}
The \ac{atp} was developed as a general protocol to underpin social networks and forms the basis of  \bsky.
The protocol defines high-level interactions between components of distributed social applications (\eg reading data from a user repository).
\ac{atp} is easily extensible and does not define the exact exchanged data types.
Those types are defined in \emph{lexicons} that can be created by the community and are organized in namespaces identified by \ac{dns}-like names.
For example, the lexicon \texttt{app.bsky} defines the type \texttt{app.bsky.feed.post}, which corresponds to a social media post in the \bsky application.

\begin{sloppypar}
\para{Decentralized Identities}
Users in \bsky are identified via a unique \ac{did}~\cite{did}, \eg, \texttt{did:plc:ewvi7nxzyoun6zhxrhs64oiz}.
A \ac{did} is immutable and identifies a user uniquely in the network.
This enables users to migrate between different servers hosting their data while maintaining their social graph.
\acused{diddoc}
\end{sloppypar}

\begin{sloppypar}
Each \ac{did} points to its associated \emph{\ac{diddoc}},
a document that stores service information about the user. This includes the endpoint of the \ac{pds} storing the user's data \emph{repository}, the user's handle, as well as public keys to verify signatures on user content.
There are currently two supported \ac{did} schemata: \emph{PLC} and \emph{WEB}.
They differ in how the \ac{diddoc} are retrieved:
\begin{enumerate*}
	\item for PLC \acp{did}, the associated document is downloaded from the \texttt{plc.directory} service, which is operated by \bskyCo.
	\item WEB \acp{did} consist of a \ac{fqdn}. The associated document must be located at \texttt{https://<fqdn>/.well-known/did.json}
\end{enumerate*}
.
\end{sloppypar}

\para{User Handles}
\label{sec:background:components:handles} 
To provide human-friendly identification, users are addressable via mutable \emph{handles}.
Handles are stored in the DID document, linking mutable and immutable identifiers.
Handles are a \ac{fqdn} and utilize \ac{dns} for proof of ownership.

\begin{sloppypar}
There are two mechanisms to prove ownership of a handle, \eg \texttt{@example.com}:
\begin{enumerate*}
	\item via a \ac{dns} \texttt{TXT} record located at \texttt{\_atproto.example.com}, containing the \ac{did} of the user, or
	\item via a file located at \texttt{https://example.com/.well-known/atproto-did}
\end{enumerate*}
.
Both mechanisms ensure the owner of the handle also has ownership of the associated domain, enabling data ownership and user identity verification.
By default, \bsky automatically manages user keys and creates a handle in the form of \texttt{@username.bsky.social}, hiding the complexity of decentralized identities.
\end{sloppypar}

\begin{sloppypar}
\para{User Data Repositories}
Repositories, or \emph{repos}, are the main data structure that stores user data.
The repos constitute a key-value store of \emph{records} and contain users' posts, likes, follows, blocks, \etc
All \bsky-related records are encoded as \ac{cbor}, as defined by the \texttt{com.atproto} and \texttt{app.bsky} lexicons.
Updates to a user's repo are signed using one of the keys contained in their \ac{diddoc}, via repo \emph{commits}.
Entries in repositories can be identified uniquely within the network via a \acp{uri}, of the form \texttt{at://<did>/<key>},
\eg, \texttt{at://<did>/app.bsky.feed.post/3kdgeujwlq32y}, where the last component of the path marks a unique ID to distinguish repeatable records.
\end{sloppypar}

\para{Personal Data Servers}
Repos are hosted by \acp{pds}.
A \ac{pds} can hold multiple repositories, but each repository is served by just one \ac{pds} that the user can choose.
\bskyCo operates the default \acp{pds}.
However, it has recently become possible to self-host \acp{pds} and federate with the network.\footnote{\url{https://docs.bsky.app/blog/self-host-federation}}
Thus, users can migrate to different \acp{pds} while maintaining their social graph.
This requires updating the endpoint in the user's \ac{diddoc}.
The \acp{pds} also store user preferences, which are only accessible by the authenticated user.

\para{The Relay}
Checking for repo updates individually is resource-intensive, as each client would need to contact many different \ac{pds}es.
To streamline the process, a centralized \emph{Relay} aggregates user interactions across \acp{pds}.
This is a central store that replicates the repo data structures from all known \acp{pds}.
The relay then provides the \emph{Firehose} --- a real-time feed of all activity in the network that anyone can subscribe to.
The Firehose has a retention time of three days and includes user repo updates as well as informational and service-related events (\eg updates to user handles).
\bskyCo runs the default Relay and Firehose at \texttt{bsky.network}.
However, other providers could offer a competing service if they wish.

\para{Feed Generators}
A major challenge with existing social networks is the use of (opaque) commercial algorithms to generate the feeds, selecting which posts to display to which user.
In contrast, \bsky implements an open ecosystem for content recommendation. 
Any user can run their own Feed Generator that other users can subscribe to.
When creating a Feed Generator, its creators add a special record in their repository, which points to the data dissemination endpoint for the feed.
The Feed Generator then consumes the Firehose and produces a bespoke feed, consisting of \acp{uri} pointing to chosen posts.
Feed generators can be self-hosted, or created using one of multiple feed-generator-as-a-service.

\para{Labelers}
To support feed generation and other content classification activities (\eg filtering hate speech), \bsky introduces the concept of \emph{Labelers}. 
Labelers are services that attach labels to objects in the network.
Users can then leverage these labels to filter content.
In practice, these labels are a simple short string (\eg \texttt{nsfw}).
Some labels are predefined and have hardcoded behavior in other components,
such as \texttt{!hide}, which hides the content without an option to click through.
Others are \emph{custom} labels, with custom behaviors.
For example, a label might indicate that a warning message should be placed over the content.
Note, labels can also be rescinded by a Labeler publishing the same label for the same target with the addition of a negation mark.

Labelers operate as regular accounts with a repository for their activity.
Each Labeler publishes a service information record in its repository, describing the values and default actions to be taken for its labels, \eg to display a warning.
Functionally, a labeling service implements an endpoint that publishes the labels.
The endpoint is listed in the \ac{diddoc} of the hosting account and is publicly accessible without authentication.
Any entity can connect to this endpoint to retrieve the stream of labels produced.
Recently, the ecosystem opened enabling anyone to run a Labeler.\footnote{\url{https://docs.bsky.app/docs/advanced-guides/moderation}}

\para{User Preferences}
The goal of the moderation system is to give each individual the flexibility to decide how they interpret and action the labels.
\bsky allows users to determine their moderation policy by describing their preferences, in terms of which labels should trigger which actions. 
The preferences are a non-public setting defining Labelers the user subscribes to and reactions to labels produced by them.
For each Labeler and label, the users can choose to ignore it, show a warning, or have the content hidden entirely.

\para{The AppView}
The \emph{AppView} collates the data produced across the network into a usable format, and makes it available to clients via a public API.
In practice, the AppView consumes the Firehose and information from other system components, stores them in a database, and provides the results in an easy way for a final app to display.
There is currently one \bsky AppView, operated by \bskyCo.

\para{The Client}
Client applications provide a usable frontend to the network.
They communicate with a user's \ac{pds} and the AppView to build the timeline displayed to users.
Note, \bsky does not mandate a single client implementation.

\section{Datasets}\label{sec:datasets}
\para{User Identifier Dataset}
Each user in \bsky is identified via a unique \ac{did}.
Utilizing the \texttt{sync.listRepos} call offered by the \bsky Relay, we obtain a list of all active \bsky users and their \acp{did}.
This additionally returns each user's latest respective repo commit version.
We query the endpoint weekly to learn of changed repositories during March and April 2024, obtaining a total of $5{,}591{,}824$ identifiers.

\para{DID Documents and FQDN Handles}
We then download the \acp{diddoc} (if available) for all user identifiers obtained in the previous step.
Recall that these documents list fully qualified domain name (FQDN) handles, endpoints for users' \acp{pds}, and other service information.
We obtain documents both from the centralized \texttt{plc.directory} service,\footnote{\url{https://plc.directory/}} operated by \bskyCo, used as the default for account creations (the \texttt{did:plc} method), 
as well as \texttt{.well-known/did.json} paths via HTTPs (the \texttt{did:web} method) (see \Cref{sec:background}).
We download the full snapshot over the course of one week in March 2024.
This yields a total of 5,077,159 \acp{diddoc} from the PLC server, with an additional six using the \texttt{did:web} method.
We extract FQDN handles from the \acp{diddoc} for further analysis.

\para{Repositories Dataset}
We download a snapshot of all users' repositories on April \nth{24} containing the public actions of users (\eg posts, likes, blocks, \etc).
We do this for the complete list of user identities and corresponding repository versions from the User Identifier Dataset.
We utilize the \texttt{sync.getRepo} endpoint of the Relay service to download a copy of each repository.
Since the Relay constantly crawls \acp{pds} and caches repositories locally, we are able to download \emph{all} repositories, irrespective of whether they are served by self-hosted \acp{pds}.
This is the recommended method of obtaining repo data, and reduces load elsewhere in the network.
We gather $5{,}523{,}919$ repositories over a period of 10 days during April 2024.

\para{Firehose Dataset}
The Firehose, provided by \bsky's Relay, offers a stream of users posts from across the network (akin to Twitter's Streaming API).
We subscribe to the Firehose offered by the Relay to obtain real-time updates from the network.
This includes user activity (\eg, posts, likes, or follows, both additions and deletions) as well as informational and service events published by the Firehose.
Since the Relay gathers all federated repositories, we receive updates from the entire visible network.
We have been continuously subscribed to the firehose from %
2024-03-06%
.
Since then, until April \nth{30}, we have collected $279{,}289{,}739$ events (\Cref{tab:firehose_event_types_overview}).
The vast majority of events are repo commits, which mark an update to the content of a user's repository.
Updates can be record creations, deletions, or replacements.
Apart from repo commits, the Firehose publishes updates to a user's handle, cache invalidation messages for DID documents,
and tombstones for deleted accounts. 

\begin{table}[ht]
	\centering
	\caption{Overview of Firehose event types.}
	\label{tab:firehose_event_types_overview}
	{\small%
\begin{tabular}{lrr}
  \toprule
Event Type & \# Total & Share (\%) \\ 
  \midrule
Repo Commit & $278{,}677{,}401$ & $99.78$ \\ 
  Identity Update & $531{,}295$ & $0.19$ \\ 
  User Handle Update & $44{,}456$ & $0.02$ \\ 
  Repo Tombstone & $36{,}587$ & $0.01$ \\ 
   \bottomrule
\end{tabular}
\unskip%
	}
\end{table}

\para{Feed Generators Dataset}
Feed Generators form the basis of content curation and moderation in \bsky~(see \Cref{sec:background}).
We compile a complete list of all Feed Generators operating in the network.
Since they can be identified via records in the repositories, we utilize the downloaded repositories as well as real-time updates from the Firehose to obtain a list of all Feed Generator identifiers.
For each Feed Generator, we determine the \ac{did} of the service responsible for \emph{hosting} it, as well as the associated endpoint.
In total, we discover 43,063 Feed Generators.
We download Feed Generator metadata (\eg descriptions and information about the creators) via the \texttt{getFeedGenerator} method of the AppView.

\para{Feed Post Dataset}
We collect \acp{uri} of all Feed Generator posts from the Appview \texttt{getFeed} endpoint bi-weekly.
We then correlate the \acp{uri} with the posts in the \emph{Repositories Dataset} to obtain the full content of each post.
Although 100\% of Feed Generators with metadata were marked as both online and valid, we were only able to get posts for 93\% of them.
While there is a way to delete Feed Generator records from the repos, provided this record is still there, there is no way to distinguish between permanent and temporary unavailability. 
We thus exclude 4.3\% of Feed Generators without metadata from our analysis.

From 2024-04-16 to 2024-05-10, we collect 21,520,083 posts from 40,398 Feed Generators.
A challenge for our data collection is that Feed Generators have different policies regarding their retention of historical posts.
While some provide all the historical posts, others impose limits based on duration (\eg only posts no older than 10 days) or the number of posts (\eg only the last 100 posts).
As a result, we are not able to collect all the Feed Generator posts from before our measurement period. 
\para{Labeling Services}
To become a labeling service, an account creates a service information record in their repository and a service endpoint entry in their \ac{diddoc}.
We utilize this to compile a comprehensive list of Labelers using the repository data and real-time updates from the Firehose.

As of %
2024-05-01%
, we identify %
$62$%
 unique labeling services.
For each, we subscribe to the public endpoint listed in their \ac{diddoc} and receive a stream of labels produced by the Labeler.
We collect \emph{all} labels produced, including ones emitted before our collection period, by consuming the entire stream of labels produced by each Labeler.
If a Labeler's endpoint is temporarily unavailable,
we backfill any missed updates as soon as the endpoint is functional again.
This includes rescinded labels, as indicated by a negation of a previously emitted label.

We attempt to reconnect to the service endpoints on a daily basis, but find only $46$ labeling services functional, of which %
$36$%
 issued at least one label.
Overall, as of May 2024, we collect %
$3{,}402{,}009$%
 interactions from labeling services, including %
$23{,}394$%
 rescinded labels.

\section{User Activity}\label{sec:core}
First, we analyze the user activity on \bsky.
We study the growth dynamics of the platform, the most popular accounts, and its current size compared to other, established social networks.
This allows us to understand the platform's current status and its user base.
We deliberately omit a deeper analysis of the social graph, as \bsky does not introduce novel solutions in this regard.
Finally, we investigate whether the theoretically flexible \ac{atp} infrastructure is used to distribute content not related to the social network itself.

The repositories dataset reveals an accumulation of operations (\eg follows, likes) through which \bsky users have interacted since the creation of the platform in Nov 2022.
We observe a total of 740M likes, 225M posts, 160.9M follows, 77.9M reposts, and 10.8M user blocking operations.
In this section, we analyze the platform's growth, its current status and language-specific communities.

\begin{figure}[t]
	\centering%
	\includegraphics[width=\linewidth]{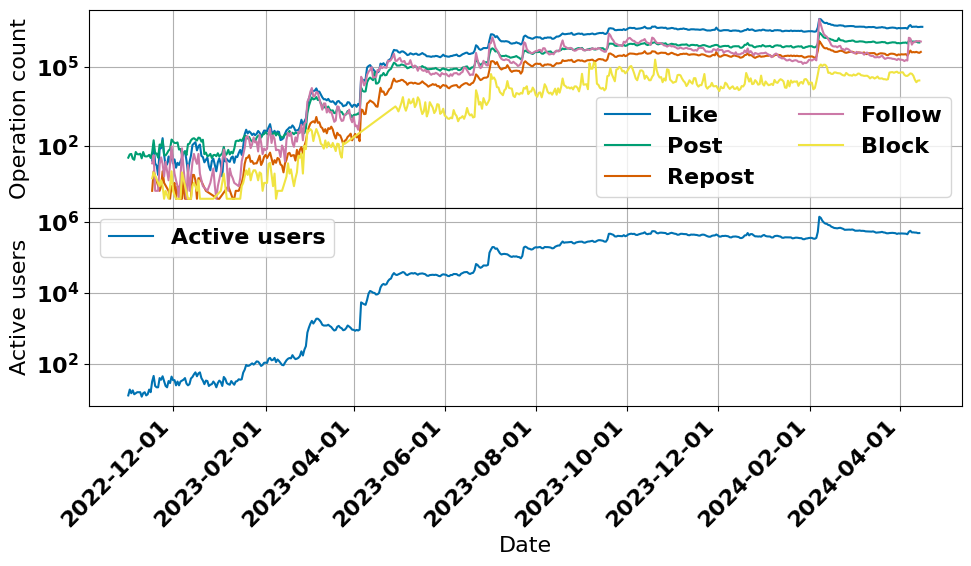}
	\vspace{-.7cm}%
	\caption{Daily operation and active user counts. }
	\label{fig:operationAndUserActivity}
\end{figure}

\para{Growth}
We observe the initial activity in users' repositories dating back to November 2022, when Bluesky was launched (\Cref{fig:operationAndUserActivity}).
This coincides with Elon Musk's acquisition of Twitter and the subsequent firing of half of the employees. 
The events following the takeover spurred interest in decentralized social media platforms~\cite{he2023flocking}.
\bsky experienced a large growth in daily user activity over the subsequent eight months following November 2022.
The number of active users increased from mere hundreds in December 2022 to hundreds of thousands by July 2023.
Another significant surge in daily user activity occurred in February 2024, when \bsky transitioned from an invite-only platform to a public one.
While \bsky experienced initial growth, we also observe stagnation and even a decrease in daily active users.
The number of daily active users decreased by $\approx$ 60K between March and May 2024.

\begin{figure}[!ht]
	\centering%
	\includegraphics[width=\linewidth]{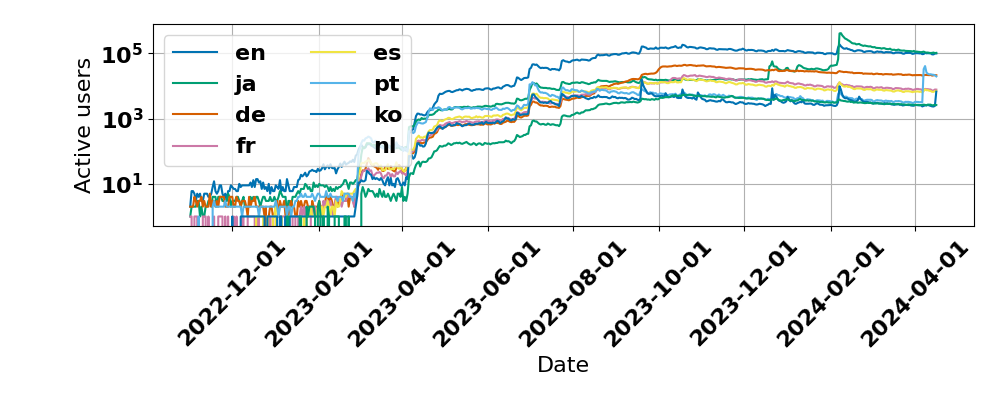}
	\vspace{-.7cm}%
	\caption{Active user counts of individual language-specific communities.}
	\label{fig:langActivity}
\end{figure}

We further investigate the activity of $\approx$ 2M \bsky users, who posted at least once, based on the self-assigned language tags attached to their posts (\Cref{fig:langActivity}).
We verify a small ($\approx 0.1\%$) random portion of these posts and find that they indeed correspond to the language indicated in the tags.
Language-specific communities roughly follow the global trend of user activity but we also notice some discrepancies.
For instance, opening the platform to the public in February 2024 significantly increased the number of active Japanese-speaking users while the German-speaking community remained largely unaffected.
In April, the Portuguese-speaking community experienced a sharp growth from $\approx$ 3K to $\approx$ 30K, most likely caused by direct marketing actions and recent documentation translation~\cite{bsky_blog_pt}.

\para{Current Status}
As of April 2024, we observe a consistent presence of around 500K active users daily on the platform, contributing approximately 3M likes, 800K posts, and 300K reposts daily (\Cref{fig:operationAndUserActivity}).
The network size is comparable to Mastodon but significantly smaller than Twitter with 1M~\cite{mastodon_users} and 245M~\cite{twitter_users} daily active users, respectively.
While English remains the platform's main language ($\approx$ 800K users), more than 700K users use Japanese language tags, suggesting a diverse user base.
The platform has recognized the importance of the community, for instance, by implementing a dedicated \emph{kawaii} mode.\footnote{\url{https://bsky.app?kawaii=true}}
Additionally, Portuguese and German emerged as the next popular languages among users.

\para{Account popularity}
We observe the account popularity based on likes and block operations.
The most popular account is the \bsky official account (775K followers as of April 2024), which posts updates on the platform.
The other popular accounts belong to popular American newspapers and independent journalists, such as the Washington Post and the NY Times, each with over 200K followers. On the other hand, the most blocked accounts belong to celebrity impersonators and propagandists. For example, the most blocked account impersonates Jordan B. Peterson, while the next most blocked account is an anti-vaccine propagandist.
Both accounts received $\approx$ 15K blocks.

\para{Non-\bsky content}
We find that the Firehose also distributes content \emph{not} covered by the \bsky-related lexicons.%
These are records in user repositories targeted for a different AppView, \ie, \bsky cannot decode or display them.
We find $1{,}855$ events (out of $\approx 280$M) relating to these records.
Among the targeted applications, we find WhiteWind\footnote{\url{https://whtwnd.com/}} to be the most popular public application.
WhiteWind attempts to bring long-form blogging to \ac{atp}.
Users can log in using their \bsky account and write articles using markdown.
The articles are then saved in the user's repo, hosted on their \ac{pds}.
The WhiteWind AppView and Frontend can decode the entries and display them on the website.

These alternative applications require the \bsky infrastructure (\eg the Firehose) to index and re-publish non-\bsky content.
The ability to use already deployed infrastructure without additional cost is beneficial to the growth of the ecosystem.
Due to the current limited size of those applications, they have a limited impact on the infrastructure.
However, it remains to be seen whether this remains viable as those platforms grow.

\para{Takeaways}
\bsky has experienced significant growth since its launch in November 2022.
However, the growth has been driven mostly by specific events such as the public launch and stagnates between those events.
\bsky is still significantly smaller than its centralized counterpart and does not yet show signs of the snowball effect or mass user migration.
This is further confirmed by the current lack of popular celebrity or institutional accounts.
At the same time, the good user experience and a rapidly increasing number of features could enable \bsky to attract specific country- or interest-specific communities.
This is exemplified by a recent mass migration of Brazilian users after X/Twitter was banned in the country~\cite{cnbc}.\footnote{Unfortunately, the migration happened outside of our data collection period.}
Finally, opening the platform infrastructure to third-party applications could further boost the growth and diversity of the ecosystem but for now, remain in their infancy.

\michal{Maybe mention we don't see many zombie accounts as the active users march the total number of users.}

\begin{figure}[!ht]
	\centering%
	\includegraphics[width=0.87\linewidth]{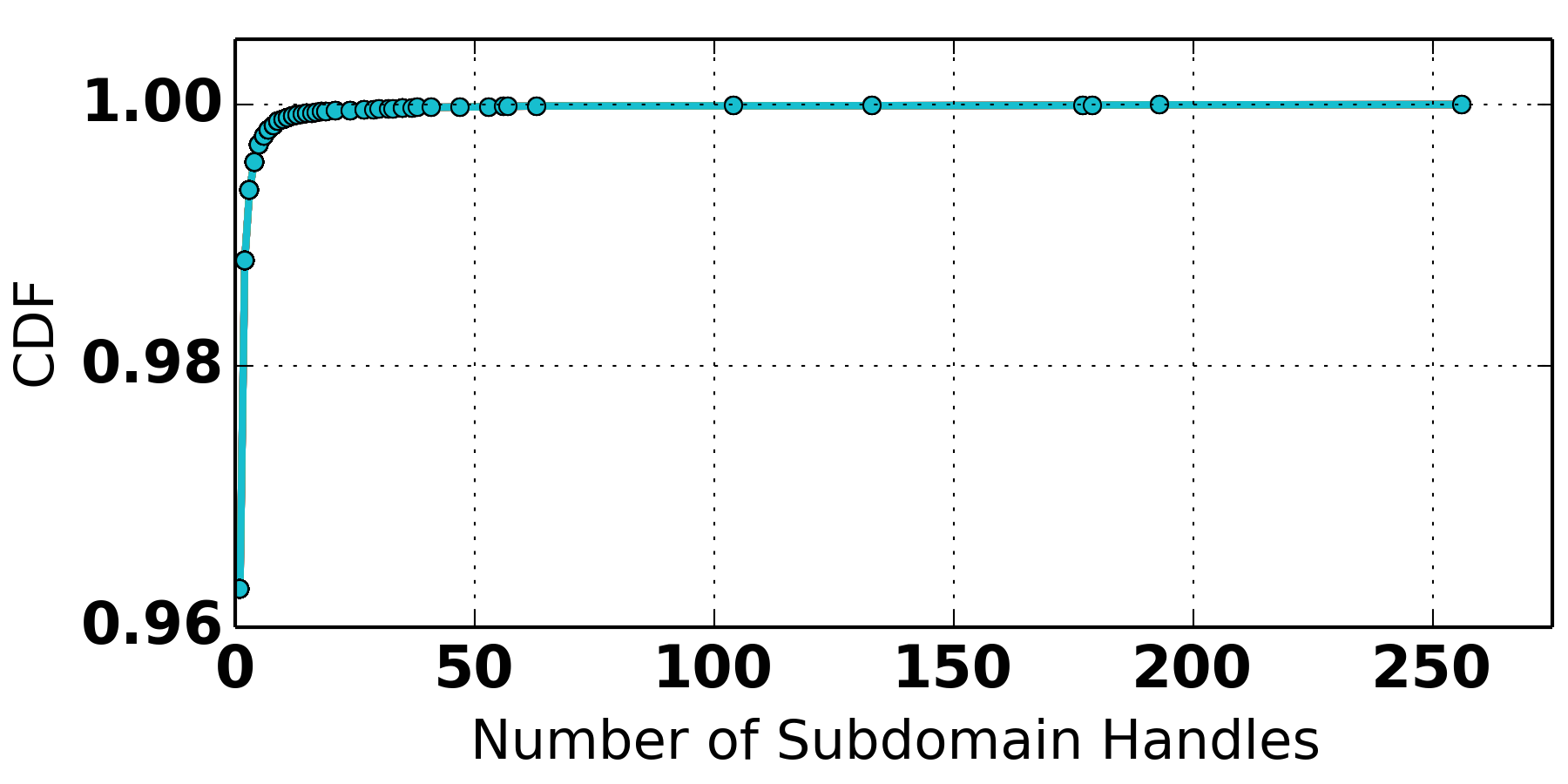}
	\caption{Number of subdomain handles per registered domain name (effective second-level domain); for clarity, we exclude subdomains of \texttt{bsky.social}, which account for 98.9\% of all observed FQDN handles.
	}
 \label{fig:cdf_handles}
\end{figure}

\section{(De)centralized Identity}
\label{sec:identity}

\bsky supports decentralized identities, allowing users to link their accounts to their domain names and manage their own keys. This frees the user from a dependence on a single identity provider.
However, to simplify the account creation process, the platform also offers a custodial identity creation that automatically creates a subdomain handle under the \texttt{bsky.social} domain.
We investigate user choices, adoption trends, and their possible implications for security and trust.

\para{Subdomain Handles Concentration}
We analyze 5,077,159 FQDN handles, sourced from DID documents, to investigate the concentration of subdomain providers.
Despite the theoretical openness of handle creation, we observe a notable concentration of subdomain name handles with 98.9\% of them under \texttt{bsky.social}.
The prevalence of such handles is expected, given their convenient management.
\bsky offers subdomain handles at no cost, making them readily accessible during the Bluesky account creation process.
Moreover, users can employ these handles without having to link them to a DID.

Next, we investigate the remaining 57,202 FQDN handles to identify alternative operators and services (\Cref{fig:cdf_handles}).
This helps us study to what extent the openness of the \bsky design translates into actual diversity.
Interestingly, no operator exceeds a few hundred FQDNs per registered name.
For instance, we find 256 FQDNs under \texttt{swifties.social}, and 179 and 133 FQDNs under \texttt{tired.io} and \texttt{vibes.cool}, respectively, both managed by Skyname.\footnote{\url{https://skyna.me}}
Those domains offer dedicated support for \bsky and facilitate the handle migration process.
We also observe generic subdomains.
For instance, we identify 35 accounts that use their \texttt{github.io} subdomain as Bluesky handles.

Using services like \texttt{bsky.social} and other dedicated subdomain providers offers the benefit of easy integration into the Bluesky ecosystem.
However, this convenience may come at the cost of giving control over data custody and management to the platform's operator.

\para{Self-Managed Domain Names}
Organizations and individuals with existing domains can leverage them to confirm the legitimacy of their Bluesky accounts.
This approach offers more control over data privacy but adds security and infrastructure management responsibilities.

We extract 51,879 registered domains (\ie, effective second-level domain names) from FQDNs using the Public Suffix List \cite{mcquistin2023first} to identify prominent brands within the ecosystem.
We cross-reference the registered domain names with the Tranco popularity list \cite{pochat2018tranco} and identify only 1,436 (2.8\%) entries within the top 1 million ranking.
These include domains associated with tech companies (\eg \texttt{amazonaws.com}, \texttt{microsoft.com}, \texttt{cloudflare.com}), media outlets (\eg \texttt{cnn .com}, \texttt{nytimes.com}, \texttt{washingtonpost.com}), and universities (\eg \texttt{stanford.edu}, \texttt{columbia.edu}).
The limited representation of domains associated with major organizations suggests limited engagement with the Bluesky platform.

\begin{sloppypar}
\para{Validating Handle Ownership} Bluesky services currently recognize DIDs derived from either \texttt{did:web}, defined by the W3C Credentials Community Group~\cite{did:web}, or \texttt{did:plc}, established by \bskyCo for Atproto~\cite{did:plc}.
Interestingly, we identify only six \texttt{did:web} identities. This observation could stem from the accessibility of the mechanism supported by \bskyCo, along with the distinction from domain name handles, particularly in the immutability of the identifier.
Unlike domain name handles, the \texttt{did:web} domains associated with a DID cannot be changed~\cite{kleppmann2024bluesky}.

With \texttt{bsky.social} (\texttt{did:plc}) domain handles, subdomains are automatically linked to their DIDs by hosting \texttt{/.well-known/atproto-did} files.
We further explore the two mechanisms employed by other FQDN handles to validate handle ownership using active measurements.
We gather 52,160 DIDs corresponding to FQDN handles beyond the \texttt{bsky.social} domain.
The vast majority of 51,497 (98.7\%) FQDN handles contain DID entries (\eg, \texttt{did=did:plc:cpa2egh7gaaesf2hqc2vuosp}) stored in the DNS \texttt{TXT} records of the \texttt{\_atproto} subdomains, while only 663 (1.3\%) are stored in the \texttt{/.well-known/atproto-did} files.
Several factors could influence the discrepancy.
Multiple online guidelines indicate step-by-step procedures for adding \texttt{TXT} records via registrar platforms without requiring in-depth knowledge of DNS.
Furthermore, storing this information in DNS avoids configuring and running a webserver just to prove domain ownership.
\end{sloppypar}

\begin{table}[t]
	\centering
 \caption{Domain name handles per registrar.}
\begin{tabular}{lrrr}
  \toprule
IANA ID & Registrar Name & \# Total & Share (\%) \\ 
  \midrule
1068 & NameCheap, Inc. & 8,252 & 20.94\% \\ 
1910 & CloudFlare, Inc. & 4,514 & 11.46\% \\
895  & Squarespace Domains  & 4,453 & 11.30\% \\
146 & GoDaddy.com, LLC & 2,835 & 7.19\% \\
1861 & Porkbun, LLC & 2,698 &6.85\% \\
69 & Tucows Domains Inc. & 2,337 & 5.93\% \\
49 & GMO Internet Group & 1,796 & 4.56\% \\
\bottomrule
\end{tabular}
\end{table}

\para{Registrar Concentration}
Next, we perform a WHOIS scan to evaluate the concentration of registered domain names among various registrars identified using IANA IDs. 
We collect WHOIS data for 47,728 (92\%) registered domain names and successfully extract the IANA ID for 39,403 (76\%) domain names.
We cannot retrieve the IANA ID for all registrars.
While ICANN-accredited registrars must display the IANA IDs for new and legacy generic top-level domains (TLDs), public WHOIS records for country-code TLD (ccTLD) operators might not always contain this information.
This could be because ccTLDs opt to provide limited data in WHOIS or because registrars are locally accredited by ccTLDs, exempting them from the requirement for ICANN-accredited registrars to have an IANA ID~\cite{bayer_eu_report}.

We find a high degree of centralization of domain handles.
While we identify 39,403 registered domains spread across 249 registrars,
50\% of the domains are registered in just four registrars. Namecheap, an ICANN-accredited registrar, accounts for 21\%.
The collaboration announced in May 2023 between \bsky and Namecheap to 
streamline the process of purchasing and linking domain names to Bluesky explains the latter.\footnote{https://bsky.social/about/blog/7-05-2023-namecheap} 
We argue that dedicated support from additional registrars will be necessary to avoid the risks of excessive registrar centralization in the future.

\para{User Handles Updates}
Finally, we analyze \emph{recent} changes in user identities via the Firehose dataset.
We observe %
$44{,}449$%
 handle changes by %
$31{,}494$%
 unique \acp{did}.
This indicates that some users changed their handles more than once during our observation period.
In these changes, we register only %
$41{,}359$%
 unique handles, indicating that some users switch back and forth between handles.
While the source handle is unavailable in the update event, we investigate the \emph{final} handles the users settled on with the updates.
We observe that $23,817$ ($75.74 \, \%$) of the final handles are under \texttt{bsky.social}, while the remaining $7,630$ ($24.26 \, \%$) are under other domains.
This might suggest that as the ecosystem develops, users are more likely to switch to custom domain names, possibly due to the increasing support from alternative services.
This perhaps highlights the flexibility of the overall architecture.

\para{Takeaways}
\bsky offers a variety of options for identity management ranging from custodial subdomain handles to self-managed domain names.
Unsurprisingly, the vast majority of users opt for the convenience of automatic subdomain handles under \texttt{bsky.social} that resembles traditional social media authentication.
This leaves \bsky in full control over user identities but also enables the platform to avoid the trap of fully decentralized, complex identity management that could deter regular users.
At the same time, institutions or tech-savvy users have the option to manage their keys and link their accounts to their domain names making their identity largely independent of the \bsky operators.
We observe that more decentralized options to identity management can attract users when made more accessible.
It remains to be seen whether such a flexible approach to identity management will enable the onboarding of a large number of users with the centralized approach while later allowing them to take more control over their identity as the decentralized identity solutions mature and develop user-friendly interfaces.

\section{Content Moderation}
\label{sec:labelers}
Content moderation is a crucial aspect of any social media platform.
The traditional social media platforms have faced criticism for their opaque and inconsistent moderation practices.
\bsky presents a novel approach that relies on two main elements:
\begin{enumerate*}
	\item \emph{Labelers}, which produce short textual labels attached to objects, and
	\item User preferences indicating
	\begin{enumerate*}
		\item which Labelers to subscribe to (\ie use labels from) and
		\item how to \emph{react} to a given label (\eg by hiding the post).
	\end{enumerate*}
\end{enumerate*}
In this section, we focus on the Labelers producing the labels, and later inspect the labels themselves and the process of their issuance.
We try to reveal how the moderation process is split between the official \bsky Labeler and community-run Labelers, and whether there is a significant overlap between them.
Note, the user preferences are not publicly visible and we make no attempt to reveal them due to ethical considerations.

\subsection{Labelers}
\begin{figure}[!ht]
	\centering%
	\includegraphics[width=\linewidth]{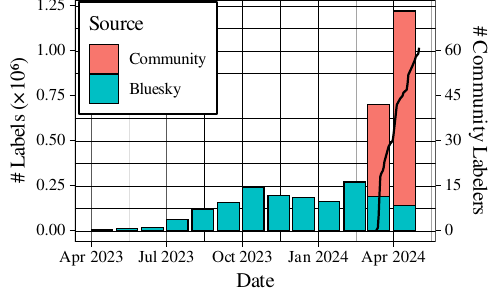}
	\vspace{-.7cm}%
	\caption{
		Number of labels produced by source per month (left-hand axis), and number of community-run labeler services over time (right-hand axis).
	}
	\label{fig:labelers_num_labels_produced_by_source_with_num_labelers}
\end{figure}

\para{Growth} 
\Cref{fig:labelers_num_labels_produced_by_source_with_num_labelers} plots the number of Labelers over time.
In April 2023, \bsky launched its first official Labeler which remained the only one for 11 months.
On March \nth{15} 2024, the platform opened up to community-driven Labelers, leading to a rapid increase in the number of Labelers.
The new Labelers produce an increasing number of labels and in April 2024 issued
$1{,}082{,}207$%
 (%
$88.7 \, \%$%
) of all the labels.

We make two observations.
\begin{enumerate*}
    \item In the federated architecture of \bsky, the centralized AppView component subscribes to all known Labelers and needs to store \emph{all} labels.
    This approach makes it relatively easy to run a Labeler.
    The required bandwidth is low and the required computing capacity depends on the implemented algorithm.
    \item Conversely, running an AppView becomes ever-more resource demanding, the more Labelers are active.
\end{enumerate*}
It remains to be seen whether this approach is scalable in the long term.

\para{Current State}
As of %
2024-05-01%
, %
$62$%
 unique accounts announced themselves as Labelers.
However, only %
$36$%
 issued at least one label.
\Cref{tab:labelers_top_5_community} lists the top 5 community-run Labelers by the number of applied labels.
Some are transparent and their authors are publicly known.
They generally post about their implementation, technical details, and challenges.
For instance, the \nth{5} most active community Labeler caters specifically to the Japanese community of  Final Fantasy 14 game players.
This service is intended to prevent accidental exposure to spoilers about new game content.
In an accompanying blog post,\footnote{\url{https://blog.usounds.work/posts/bluesky-ff14-labeler}} the author describes the implementation and notes challenges with false positives.
In other cases, the operators remain anonymous and do not provide details about their service (\eg the \enquote{AI Imagery Labeler}, which is operated by multiple anonymous individuals following a \enquote{Moderator Handbook}).\footnote{\url{https://trail-buckthorn-014.notion.site/Bluesky-AI-Imagery-Labeler-Moderator-Handbook-65ef75d92a0e4faab0ad2925fc35c85c}}

We analyze the IP addresses of the Labelers.
Most (%
$40$%
, or %
$65 \, \%$%
 of all) services run on cloud-based hosting infrastructure or are reverse-proxied.
However, we find that %
$6$%
 (%
$10 \, \%$%
) are operated from ISP-assigned residential addresses.
The remaining %
$16$%
 (%
$26 \, \%$%
) were not functional and no endpoint could be determined.

\begin{table*}[t]
	\centering
	\caption{Top 5 community labelers by number of labels applied.}
	\label{tab:labelers_top_5_community}
	\small
	\begin{tabular}{rrlrlp{5.5cm}}
		\toprule
		Rank & \# Applied & Name & Likes & Operator & Description \\
		\midrule
		1 & $1{,}360{,}224$ & Bad Accessibility / Alt Text Labeler & $99$ & \texttt{@baatl.bsky.social} & Labels posts for missing/invalid alt text. \\ %
		2 & $76{,}599$      & XBlock Screenshot Labeler & $301$ & \texttt{@aendra.com} & Uses a machine-learning model to classify screenshots by origin.\\ %
		3 & $73{,}875$      & No GIFS Please & $88$ &  & Labels GIFs. \\ %
		4 & $56{,}517$      & AI Imagery Labeler & $546$ &  & Labels AI-related posts by hashtags. \\ %
		5 & $10{,}024$      & @ff14labeler.bsky.social & $15$ & \texttt{@usounds.work} & Labels Final Fantasy 14 content spoilers.\\ %
		\bottomrule
	\end{tabular}
\end{table*}

\subsection{Labels}

\para{Number of labeled objects}
So far, Labelers have applied labels to a total of %
$3{,}160{,}851$%
 unique objects.
Due to the dynamic development of the Labeler ecosystem, %
$1{,}122{,}226$%
 (%
$36 \, \%$%
) of those objects were labeled in April 2024 alone.
For context, in April, the entire network produced a total of %
$26{,}467{,}002$%
 posts.
Of those, %
$1{,}114{,}848$%
 (%
$4.21 \, \%$%
) were labeled with at least one label.
The most commonly labeled objects are posts (%
$99.63 \, \%$%
),
followed by entire accounts (%
$0.23 \, \%$%
),
and profile pictures/banners (%
$0.14 \, \%$%
).

Labels cause different behavior depending on the object they are applied to.
For labeled posts, the post itself, or just the media attached to it, is subject to action by the client.
Posts can be hidden, their media blurred, \etc
Profiles usually receive labels for their profile picture or banner image.
In these cases, the labels only have minor effects: content is still shown, although the profile picture or banner of the account is blurred or hidden.
Finally, labels can also be applied to entire \emph{accounts}, as identified by their \ac{did}.
Configured label behavior applies to the entire account, \eg, hides all posts from that account, if chosen by the user.

\begin{table*}[ht]
	\centering
	\caption{Label targets with most-applied labels.}
	\label{tab:labelers_label_targets_with_example_labels}
	{\small%
\begin{tabular}{lrrp{12cm}}
  \toprule
Object Type & \# Objects & Share (\%) & Top Labels \\ 
  \midrule
Post & $3{,}332{,}727$ & $99.63$ & no-alt-text ($1{,}359{,}752$), porn ($1{,}256{,}305$), sexual ($375{,}620$), ai-imagery ($56{,}603$), tenor-gif ($54{,}968$) \\ 
  Account & $7{,}601$ & $0.23$ & !takedown ($2{,}643$), spam ($1{,}067$), ai-imagery ($582$), impersonation ($575$), transphobia ($311$) \\ 
  Banner/Avatar & $4{,}706$ & $0.14$ & sexual ($2{,}538$), porn ($1{,}742$), nudity ($208$), gore ($104$), self-harm ($35$) \\ 
  Other & $121$ & $< 0.01$ & porn ($65$), sexual ($30$), !takedown ($12$), nudity ($5$), gore ($2$) \\ 
   \bottomrule
\end{tabular}
\unskip%
	}
\end{table*}

\para{Label values}
We identify a total of %
$222$%
 distinct label values.
After cleaning the data (\eg removing negations without previous applications), there are %
$196$%
 distinct label values.
We find that Labelers mostly deal with disjoint parts of the network.
Only %
$100{,}888$%
 (%
$3.2 \, \%$%
) of the labeled objects have labels by multiple services applied on them and %
$9$%
 objects are labeled with the same label by different Labelers.

Looking at \Cref{tab:labelers_label_targets_with_example_labels}, the most-applied label for posts is \texttt{no-alt-text}, applied by the most popular community Labeler.
This label marks posts with attached media missing an alternative text description for the media.
The next most frequently applied labels (\texttt{porn} and \texttt{sexual}) describe \ac{nsfw} content and are almost exclusively applied by the official \bsky Labeler.
The \texttt{tenor-gif} label is applied by a community Labeler.
It marks posts containing a reaction GIF from the popular Tenor\footnote{\url{https://tenor.com/}} GIF keyboard.

Label values prefixed with an exclamation mark are reserved and hold special meanings.
They are valid only when issued by the official \bsky Labeler.
All users are subscribed to the official \bsky Labeler and unsubscribing is not an option.
Their behavior is hardcoded in the client implementation and other system components.
As an example, the \texttt{!takedown} label causes labeled content to be purged from the network.
This label can be applied to posts, but also to entire accounts via their \ac{did}.
For the latter, it causes the entire account to be removed from system components and requests for its content to be discarded.
The labels \texttt{porn}, \texttt{sexual}, and \texttt{graphic-media} also have hardcoded behavior, but can be emitted by any Labeler.
Content labeled with them becomes inaccessible to users under the age of 18.

Another challenge is that there is no official list of potential label values to apply (apart from $7$ labels, of which some are exclusive to the official \bsky Labeler as outlined earlier).
As such, different Labelers use different labels with similar meanings.
While this can pose issues regarding coherent labeling, it also offers flexibility to developers.
Note that Labelers have to provide descriptive metadata when declaring the list of label values they emit.
Ultimately, Labelers play just one part in the overall moderation infrastructure:
users choose which Labelers to subscribe to, and how their client should react to specific labels produced by these services.
Because users choose which Labelers to subscribe to, there is a potential challenge in how to discover appropriate Labelers, and make informed decisions about which labels to trust.
\gareth{The above contains no stats, and therefore the interpretation text feels a bit fluffy. Can you give some estimates of the overlaps}

In general, we find that label values generally show little overlap between the \bsky Labeler and community Labelers, \ie, they seem to deal with mostly disjoint topics:
Only %
$56{,}856$%
 (%
$1.8 \, \%$%
) objects are labeled by the \bsky Labeler and any community Labeler.
The \bsky Labeler mostly labels \ac{nsfw} content and upholds some community standards whereas the community Labelers seem to operate in specific niches.
This is enabled by the flexibility (or lack of predefined) label values and user choice in how to react to labels.

\subsection{Label issuance}
Finally, we gather insights about the process of issuing labels. 
\begin{figure}[!ht]
	\centering%
	\includegraphics[width=\linewidth]{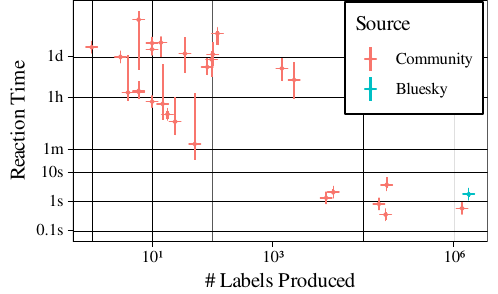}
	\vspace{-.7cm}%
	\caption{Number of labels produced by source vs. reaction time --- median and quartiles shown.}
	\label{fig:labelers_labels_per_source_and_reaction_time_scatterplot}
\end{figure}
We first analyze how long it takes for a label to be issued.
\Cref{fig:labelers_labels_per_source_and_reaction_time_scatterplot} shows
the reaction time (median and \nth{1}/\nth{3} quartile) and the number of labels produced per Labeler.
To calculate the reaction time, we exclusively look at new \emph{posts} received from the Firehose since %
2024-03-06%
.
We do this to avoid other objects which retrieve labels less frequently (\eg accounts), and labels applied retroactively to older posts.
The more labels a Labeler produces, the faster they are generally in reacting to new posts, suggesting a high degree of automation.
This is reinforced by the variability in reaction times.
Labelers producing fewer objects are generally more variable in their reaction time, indicative of a manual process rather than an automated one.
Note that the Labeler with the most labels in total is the one operated by Bluesky, which has been running for $\approx 11$ months more than the other ones.

\begin{figure}[!ht]
	\centering%
	\includegraphics[width=\linewidth]{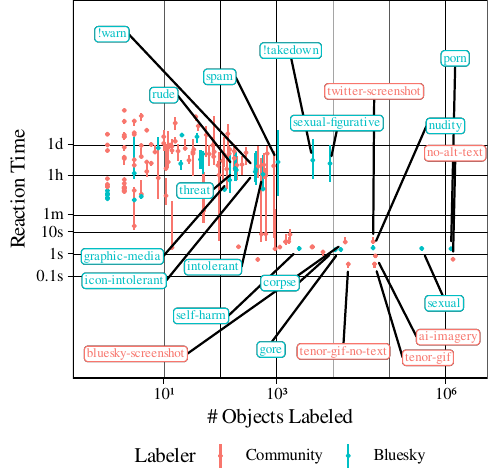}
	\vspace{-.7cm}%
	\caption{Number of labels produced by source vs. reaction time --- median and quartiles shown.}
	\label{fig:labelers_label_values_per_source_and_reaction_time_scatterplot}
\end{figure}

In \Cref{fig:labelers_label_values_per_source_and_reaction_time_scatterplot} we investigate the number of objects labeled per label \emph{value} (\eg, \texttt{porn}) and the labeler's reaction time.
The figure shows the median and \nth{1}/\nth{3} quartile and the color indicates the producing Labeler.
On the lower right-hand side, we find the most-active community Labelers, applying \eg \texttt{no-alt-text} and \texttt{ai-imagery}, as well as the screenshot-classifying Labeler.
Reaction time for these is generally very low (\eg $< 10\,\text{s}$), as the systems are likely automated.
The labels in the upper left-hand corner are applied rarely and are mostly produced by community Labelers.
We find that some of these are simply experiments by early adopters, while the majority are probably applied manually.

For the official \bsky Labeler, we observe two groups of labels:
In the lower half of the plot, we find \texttt{porn}, \texttt{nudity}, \texttt{corpse} \etc, which are applied within seconds, indicating automated systems.
On the other hand, labels such as,\eg, \texttt{spam}, \texttt{sexual-figurative}, \texttt{intolerant}, and \texttt{!takedown} take longer to be applied, pointing to manual processes.
It seems that heavy-handed moderation decisions such as removing data are deliberated instead of automated.
This is reassuring, especially for the \texttt{!takedown} label having significant consequences for the affected users.

\para{Takeaways}
\bsky combines mandatory\footnote{We note that, theoretically, one could run an alternative AppView component that would ignore labels issued by \bskyCo.} platform-run moderation with a plethora of custom, user-led Labelers.%
The open Labeler ecosystem is still in its infancy and mostly issues specific labels (\eg GIFs from a specific platform) avoiding more controversial topics (\eg fake news or hate speech).
Notably, the ease of running a Labeler and the flexibility of the system already enable it to be used not only for specific, niche topics, but also for downstream content recommendation (\cf \Cref{sec:feed_generators}).
While we observe a high degree of automation in the label issuance process, some labels are still likely applied manually.
This is especially true for the more subtle labels, such as \texttt{threat} or \texttt{intolerant}.
We expect more labels to be issued automatically in the future as the content rate increases and technology matures.

As \bsky grows, the platform might be also exposed to increasing regulatory pressure to moderate content.\footnote{\cf \url{https://bsky.app/profile/aaron.bsky.team/post/3l3gerugkbt27} for a recent example of increased need for moderation.}
This creates the risk of running forced, centralized moderation before the content is distributed to the network and limiting the role of decentralized Labelers.
We note that it is already possible for \bskyCo to perform \enquote{infrastructure takedowns}, instantly removing data from their infrastructure in cases of clearly illegal content.
While running a Labeler is relatively easy, the incentives for their operators to scale up their operations are unclear.
In the long run, the decentralized Labelers might thus cover the issuance of high-quality labels for niche use cases, while the centralized moderation would handle the bulk of the sensitive content (\eg CSAM) and remove it from the network.

\section{Content Recommendation}\label{sec:feed_generators}
\bsky allows users to personalize the content displayed on their timelines by subscribing to Feed Generators. 
Users can subscribe to multiple feeds and switch between them seamlessly.
We analyze the Feed Generators' ecosystem, their impact on the network and the platforms used to run them.

\subsection{Feed Generators}
\begin{figure}[!ht]
	\centering%
	\includegraphics[width=\linewidth]{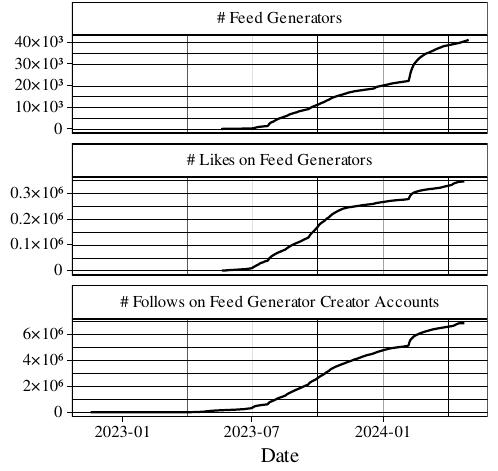}
	\vspace{-.7cm}%
	\caption{Cumulative sum of number of Feed Generators, likes on Feed Generators, and followers of Feed Generator creator accounts, over time.}
	\label{fig:feed_generators_cumsum_fg_with_likes_and_follows}
\end{figure}

\para{Growth}
The Feed Generators were introduced in May 2023.
Since then, their number has been steadily increasing (\Cref{fig:feed_generators_cumsum_fg_with_likes_and_follows}).
We also observe a significant increase in the number of likes on Feed Generators and followers of Feed Generator creators.
As the Feed Generator subscriptions are not public, the number of likes they receive sheds light on their popularity.
Additionally, creators responsible for Feed Generators seem to be attracting more followers.
This suggests that creating Feed Generators might be a way for users to reach a wider audience. 
Similar to other Bluesky components, we observe a growth increase for all those metrics in February 2024, when the platform was opened to public use.

Our investigation shows that most of the generators provide historical data from 1 to 7 days back.
As a result, we acquire a complete list of posts only for the period between April and March 2024 when we performed our measurements.
We thus do not present historical data on the number of Feed Generator posts.

\para{Current State}
As of 2024-04-30, the network contains 40,398 reachable Feed Generators.
However, 3,782 (9.4\%) have never curated any posts, while 8,792 (21.8\%) did not curate posts in the last month and seem inactive.
Interestingly, 2,202 (0.01\%) Feed Generator posts have timestamps predating Bluesky's launch.
These include timestamps from years such as 1185, 1776 or 1923, well before even the start of the Unix timestamp. This suggests an implementation error that we reported to the Bluesky team.

Using langdetect~\cite{langdetect2021} on the Feed Generator descriptions, we detect a total of 46 languages. %
While English is the most common language (45\%), we also find a significant number of Feed Generators in Japanese (36\%), German (4.1\%), Korean (2.0\%), and French (1.9\%).
The distribution is roughly consistent with the overall language distribution on Bluesky (\Cref{fig:langActivity}).
However, we observe a lower proportion of Portuguese Feed Generators compared to the overall Portuguese-speaking user base.
This might be caused by the recent increase in Portuguese-speaking users, who might not have had enough time to create Feed Generators.

\begin{figure}[!ht]
    \centering%
    \includegraphics[width=\linewidth]{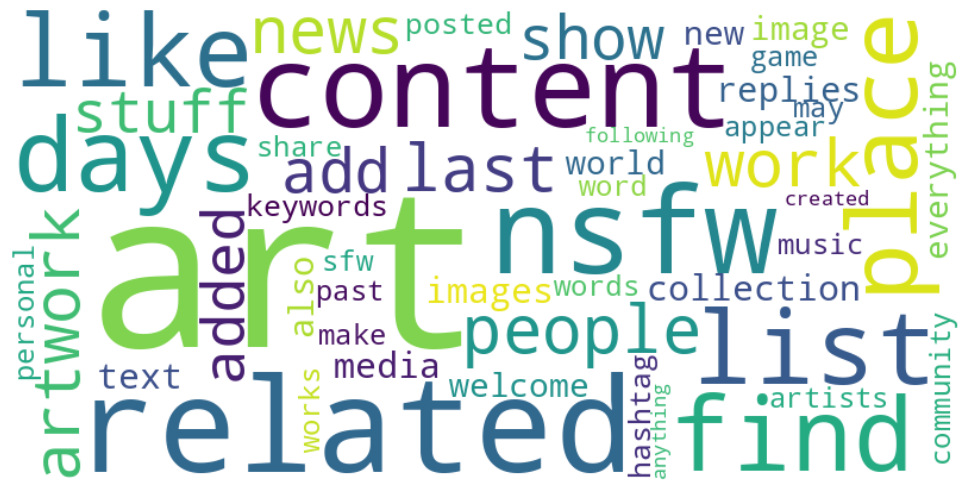}
    \vspace{-.7cm}%
    \caption{Word cloud showing the most common words found in the description of Feed Generators.}
    \label{fig:feed_generators_descritpion_word_cloud_en}
\end{figure}

We further extract the most common words present in the description (\Cref{fig:feed_generators_descritpion_word_cloud_en}).
We observe that the art community is particularly active in utilizing Feed Generators (\eg \enquote{art}, \enquote{artists}).
It provides artists with a way to showcase and potentially promote their works within the network.
This is further confirmed by links to artist content-sharing platforms (\eg tumblr, deviantart, pixiv) found in the descriptions.
Furthermore, some Labelers explicitly tag their content in the descriptions by using \enquote{nsfw} or \enquote{sfw} keywords.

\begin{figure}[!ht]
    \centering%
    \includegraphics[width=\linewidth]{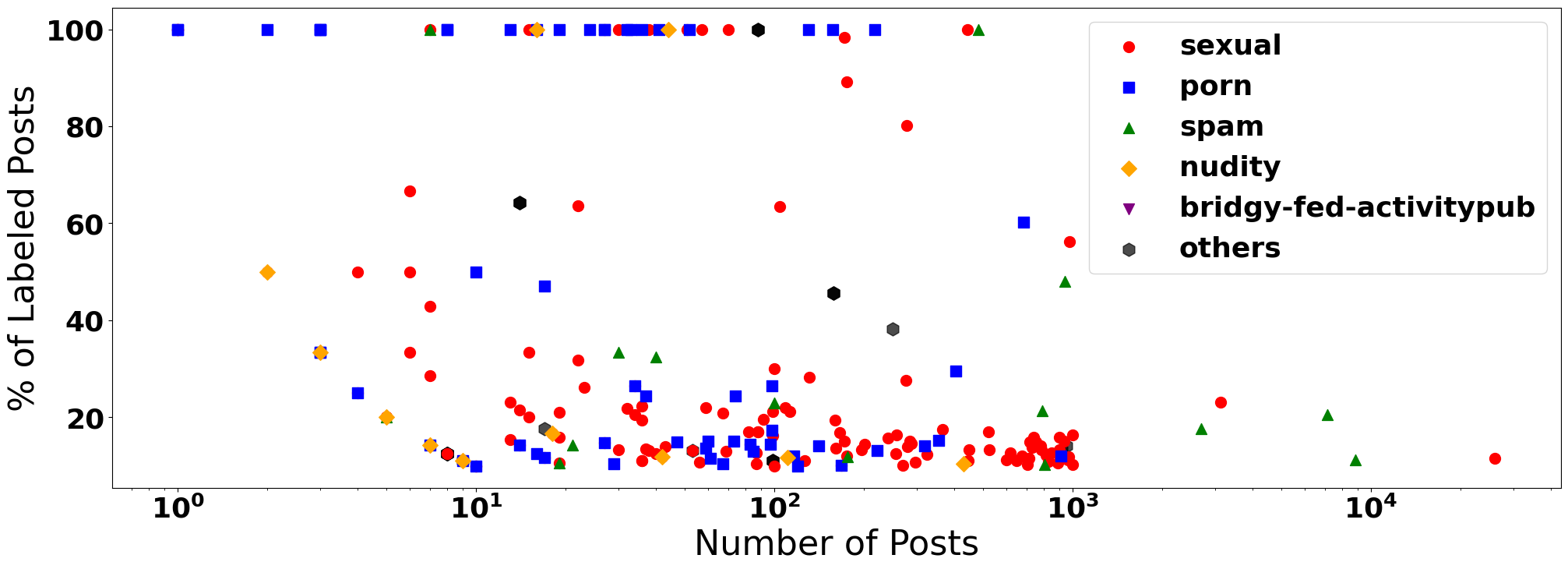}
    \vspace{-.7cm}%
    \caption{Top labels associated with posts curated by Feed Generators.}
    \label{fig:feed_generators_post_labels}
\end{figure}

We next delve deeper into the content of Feed Generators by analyzing the labels associated by Labelers with the posts.
Only 12.6\% of Feed Generators have some of their content labeled.
We focus on the 0.53\% of Feed Generators that have 10\% or more of their content labeled and report their most frequent label (\Cref{fig:feed_generators_post_labels}).
Most of these Feed Generators (0.096\%) are dedicated to explicit content (\eg \enquote{porn}, \enquote{sexual}, \enquote{nudity}) or spam.
The dominance of such labels is caused by the \bsky official Labeler that focuses on filtering this kind of content and issues the majority of all the labels.
While some independent Labelers issue labels dedicated to describing the content rather than filtering it out (\eg the Final Fantasy Labeler), such use is not yet widespread.
Importantly, the flagged content is not removed from the platform. 
Rather, each user can decide how to react to each specific label.

\begin{figure}[!ht]
    \centering%
    \includegraphics[width=\linewidth]{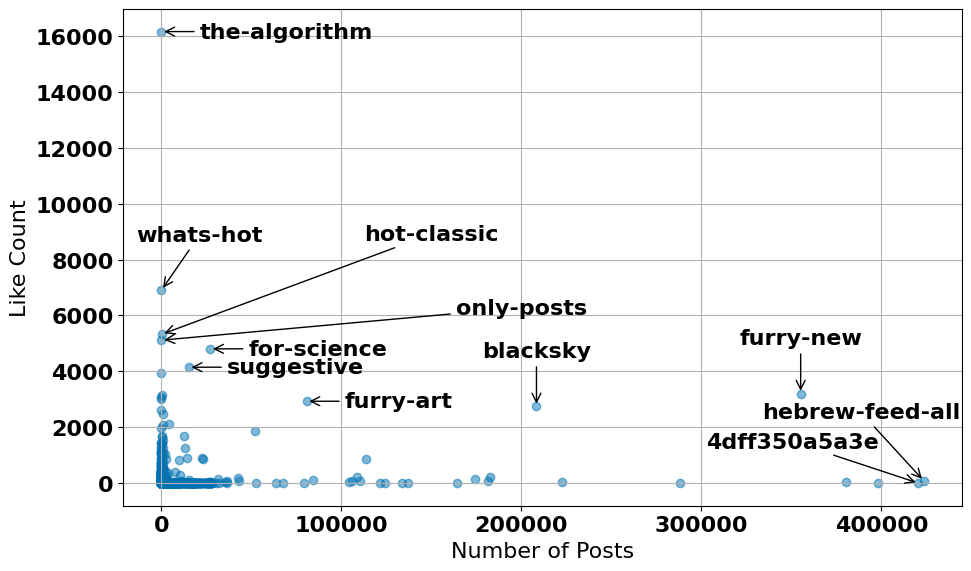}
    \vspace{-.7cm}%
    \caption{Scatter plot showing the number of posts in relation to the like count.}
    \label{fig:feed_generators_posts_vs_likes}
\end{figure}

\Cref{fig:feed_generators_posts_vs_likes} shows the number of curated posts and likes received by each Feed Generator.
We observe that the majority of Feed Generators have a low number of posts and likes.
However, a few Feed Generators curate a large number of posts ($>400,000$) or receive a high number of likes ($>16,000$).
Interestingly, the number of likes received by Feed Generators is not directly proportional to the number of posts curated by them.

To explore this, we manually investigate the Feed Generators on both extremes.
Highly-liked Feed Generators returning no posts are personalized. For instance, \enquote{the-algorithm} tailors feeds based on user likes, while \enquote{whats-hot} aggregates trending content from a user's personal network.
They do not return any content to \enquote{empty} accounts that we use for our crawls.
On the other extreme, there are automatic and aggregating Feed Generators.
For instance, \enquote{4dff350a5a3e} is a Japanese language feed tracking hundreds of thousands of posts related to the popular noodle dish \enquote{ramen}, while \enquote{hebrew-feed} automatically reposts all the content in Hebrew.
We also find active and highly-liked Feed Generators that curate their content, potentially manually.
This includes content relevant to specific communities such as \enquote{blacksky} or \enquote{furry-new}.

\begin{figure}[!ht]
	\centering%
	\includegraphics[width=\linewidth]{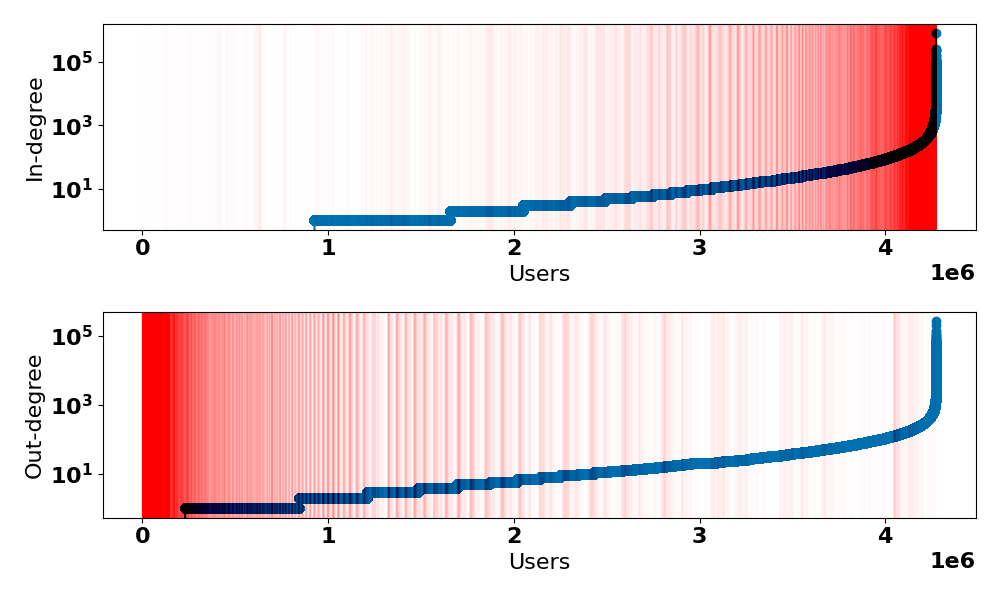}
	\vspace{-.7cm}%
	\caption{Degree distributions of users based on follow operations, with Feed Generators highlighted in red.}
	\label{fig:userDegrees}
\end{figure}

\para{Gaining Popularity}
We also investigate additional factors impacting the popularity of Feed Generators. 
\Cref{fig:userDegrees} presents the in-degree (top) and out-degree (bottom) distributions of users based on follow operations.
A red shading in both plots indicates the density of accounts that created Feed Generators.
The shade of red intensifies as the in-degrees increase and the out-degrees decrease.
This indicates that the Feed Generator accounts correspond to users with many followers.
Intuitively, popular users are more likely to create Feed Generators and Feed Generators increase the popularity of their creators.
We observe a reverse trend for the out-degree though. 
Users creating Feed Generators follow a small number of other accounts. %
We calculate the Pearson's Coefficients for various factors to estimate whether they are predictive of an account attracting followers.
We find that the \emph{number} of Feed Generators created does not correlate with the number of followers on the creator account ($r=$ %
$0.005$%
).
When calculating the sum of \emph{likes} on the Feed Generators created, however, we find correlation at $r=$ %
$0.533$%
.
This indicates that creating good Feed Generators is a way for users to gain a larger followership.

Finally, we investigate the number of feeds created per account.
A majority of users (62.1\%) manage only one Feed Generator, while 37\% of users manage between 1 and 10.
Interestingly, a small fraction (0.02\%) of accounts manage a large number of Feed Generators, exceeding 100.
An account creating the most (1799) feeds belongs to a \emph{Feed Generator As a Service} platform that simplifies feed creation.
Feeds created via this platform remain associated with the platform rather than the user account justifying the high feeds per account ratio.
It motivates us to investigate such platforms.

\subsection{Feed Generator As a Service}
\begin{figure}[!ht]
    \centering%
    \includegraphics[width=\linewidth]{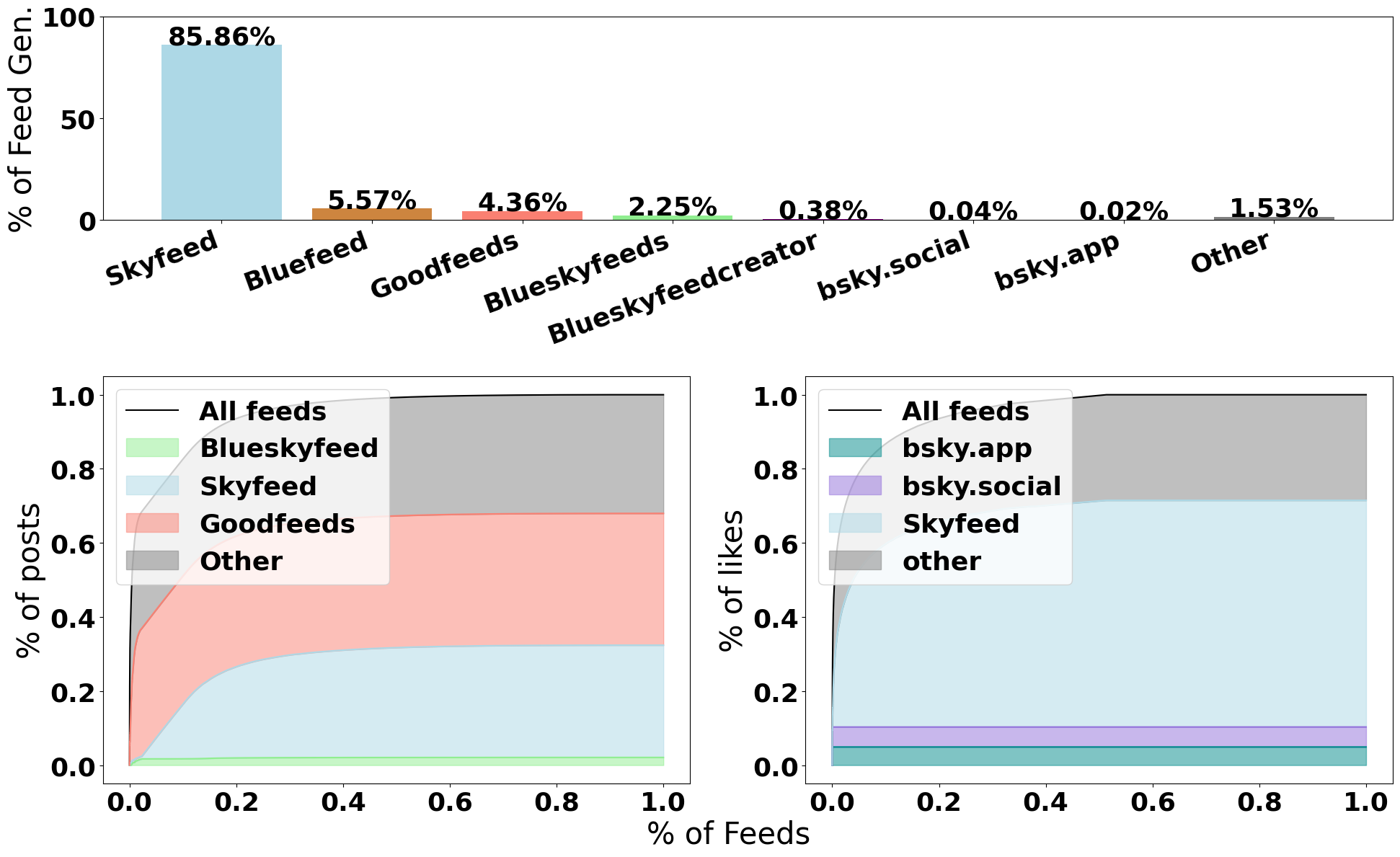}
    \vspace{-.7cm}%
     \caption{Percentage of providers hosting feed generators (top) and simplified Pareto chart of feed providers (bottom). }
    \label{fig:feed_generator_builders}
\end{figure}

We look into the Feed Generator As a Service ecosystem by analyzing the servers hosting existing Feed Generators (\Cref{fig:feed_generator_builders}).
The top three services, Skyfeed, Bluefeed, and Goodfeeds, collectively host 95.8\% of all the Feed Generators, with Skyfeed alone hosting 85.86\% of them. This may indicate another example of centralization.

The number of Feed Generators does not necessarily correlate with the number of created posts.
For instance, Skyfeed, hosting 85.86\% Feed Generators, produces only 30.3\% of the posts but accounts for 61.2\% of likes.
On the other hand, Goodfeeds, despite hosting only 4.36\% of the Feed Generators, is responsible for 35.6\% of the posts but receives 1.2\% of likes.

To better understand these differences, we analyze the features offered by each Feed Generator As a Service platform.
We provide a full comparison table in \Cref{tab:feed_comparison} in the Appendix.
The services allow their user to consume specific inputs (\eg a specific user or a feed) and filter them using labels, languages or regular expressions.

Skyfeed provides by far the most comprehensive list of features, explaining its high market share. For instance, it is the only platform that supports regular expressions. Additionally, while Skyfeed does support personalized feeds, this feature requires manual setup from the developers and is not automated. As a result, most personalized Feed Generators are currently run by their creators. Although consisting of only 0.09\% of all Feed Generators, they are among the most popular ones in the network (\Cref{fig:feed_generators_posts_vs_likes}). Adding such features to the Feed Generator As a Service platforms would require a high amount of implementation effort and increases the platform costs. However, their lack makes creating highly-quality, personalized Feed Generators challenging.

Most of the platforms offer their services for free and are run by platform enthusiasts. 
Only Blueskyfeedcreator provides paid options for feeds with additional features.
Our conversation with Feed Generator As a Service operators suggests that they try to cover their costs by donations (\eg using \url{patreon.com} or \url{ko-fi.com}) or consider running additional \bsky-related services to generate profit.
The lack of clear economic incentives puts in question the ability to scale the entire Feed Generator ecosystem.

\para{Takeaways}
In contrast to Labelers, Feed Generators already play a prominent role in the Bluesky ecosystem.
High quality feed generators are widely used by users, while automatic, spamming feed generators are less popular.
Running a popular Feed Generator is an efficient way for users to gain new followers providing a potential incentive for users to create them.
The already dynamic ecosystem of Feed Generator As a Service platforms and its simplicity of use promises a rapid growth of user-led Feed Generators.

\section{Related Work}\label{sec:related}

Several recent works have explored alternative emerging federated social networks.
Perhaps closest to our work is an initial investigation highlighting the centralization tendencies within Mastodon \cite{raman2019challenges}.
Since then, there have been several related studies of these so-called \enquote{fediverse} applications.
For instance, \cite{cava2021understanding} study the growth of Mastodon, and others have investigated how user behavior differs across server instances \cite{Binzia24,cava2022information, cava2022network}.
Additionally, some very recent works deal with \bsky from a social network perspective, which is not the focus of our work:
Quelle and Bovet~\cite{quelle2024blueskynetworktopologypolarization} investigate the social and interaction graphs for \bsky from a network-scientific perspective.
Jeong \etal~\cite{jeong2024bluetempnettemporalmultinetworkdataset} collect and investigate a dataset of user interactions with timestamp annotations to find temporal patterns.
They also investigate migration patterns from and to \bsky~\cite{jeong2024usermigrationmultiplesocial}.
One of the main challenges Bluesky attempts to overcome is decentralized content moderation, via its labeling architecture.
There have been several studies looking at content moderation labeling in alternative decentralized social networks.
For instance, Hassan et al. investigate issues with policy implementation in decentralized social networks \cite{hassan2021exploring}, and Zia et al. \cite{bin2022toxicity} proposed a federated solution to training moderation models in the fediverse.
Bluesky offers an interesting new point on the design space. To the best of our knowledge, we are the first to study this new architecture.
Beyond the above federated networks, there have also been studies of various P2P decentralized social networks, perhaps most notably Secure Scuttlebutt (SSB).
These offer a P2P event-sharing protocol and an architecture for social applications \cite{SSB}. There are also blockchain-based social networks, such as including memo.cash \cite{memo}, Steemit \cite{guidi2020socioeconomic}, and Sapien \cite{jaatun2022}. 
Finally, some work focused on assessing the centralization of decentralized platforms such as Interplanterary Filesystem (IPFS)~\cite{balduf2023cloud, wei2024eternal}.

Our work differs as we focus on an entirely different decentralized architecture: \bsky. In contrast to prior approaches, \bsky decomposes social platform functionality into a set of sub-components. These are then opened to competing providers. To the best of our knowledge, we are the first to offer a comprehensive measurement study of \bsky and its novel architecture.

\section{Discussion and conclusion}\label{sec:discussion}
We have presented the first comprehensive measurement study of the \bsky network.
The platform implements a unique, hybrid approach to federation, content moderation, and recommendation, which presents its own set of opportunities and challenges. 

\para{Ease of use and decentralization}
Fully decentralized platforms, despite their benefits, tend to be more difficult to use than centralized ones.
Manually handling cryptographic keys or server migrations is too difficult for the majority of users~\cite{whitten1999johnny}.
The \bsky approach tries to strike the right balance.
By default, the platform automatically handles key management and DNS domain creation.
While giving full control to \bskyCo, this procedure looks acceptable for most users and enables easy uptake.
At the same time, tech-savvy users can opt for more control by managing their keys and domains themselves.
Currently, only a small fraction of users (1.1\%) have chosen this option, suggesting that few people wish to exploit this opportunity, or the technical challenges remain too difficult for most to overcome.
However, the recent development of alternative services with dedicated support facilitating \bsky identity management (\eg NameCheap) might increase this number.

\para{Openness and diversity}
Openness translates into diversity when the barrier to entry is low.
\bsky has opened a number of components to the community, allowing anybody to implement competitors.
Most notably, the content moderation and recommendation sub-components enjoy a diverse and growing ecosystem.
There are currently $>$ 40k Feed Generators providing personalized feeds (\eg the-algorithm), focusing on niche communities (\eg furry-art) or serving explicit content (\eg feed-me-porn).
While \bskyCo hides some of this content from the default view, users can still access them by adjusting their settings.
Our results show that the use of content moderation is also growing.
We find $62$ Labeler accounts in total, of which $34$ are active.
Community-operated Labelers already issue the majority of labels in the network after only two months of their introduction.
The freedom to create custom labels and to re-adapt existing ones provides a high level of flexibility.
It remains to be seen whether some standardization will be necessary to avoid inconsistencies and misinterpretations as the system grows.

Importantly, \bskyCo still controls the Firehose and the AppView, which are arguably the main choke points of the system.
This enables deleting user accounts, enforcing rules, and vetting external services.
However, it is not clear how a potential migration would work, and the network effect might prevent users from switching to a new service.

\para{Scalability}
Maintaining centralized components by \bskyCo ensures a good user experience and simplifies the development of federated components.
For instance, running an independent Labeler requires only lightweight operations, while the Firehose and AppView handle the heavy lifting of aggregating a global view of the network.
However, as the platform grows, these centralized components might become bottlenecks.
Based on our measurements, we estimate that the Firehose already outputs $\approx$ 30GB of data per day per subscribed client.
In the long run, the platform might need to decentralize these components to scale further.

\para{Legal compliance}
\bskyCo introduces an inclusive approach to content moderation.
While some content is hidden by default, it is still processed and served by the platform.
This approach might be problematic if users introduce problematic content such as copyrighted material or child sexual abuse material. Furthermore, the platform uses a git-like structure for storing data.
Content can be marked as deleted, but can be still recovered from the repositories.
\lb{can we just drop this here without any proof etc.?} 
This might be problematic from the perspective of privacy-protecting laws such as GDPR.

\para{Interoperability}
\bsky is built on top of the AT protocol~\cite{kleppmann2024bluesky}, which is extensible and designed to host multiple applications.
Our analysis of the repositories shows that non-\bsky content is already present in the network,
indicating that the \ac{atp} fulfills its role as an extendable base layer to build social applications upon.
However, the platform is currently not interoperable with external social applications (\eg Mastodon) that run on another open protocol -- ActivityPub~\cite{activitypub}.
Greater interoperability could be key in increasing activity in \bsky.
Given the similar focus on openness and user portability of applications supporting the ActivityPub protocols would be a good candidate and existing bridges already point to this possibility.
We note that discussions on the integration are already ongoing in the community.\footnote{\url{https://github.com/bluesky-social/atproto/discussions/1716}}

\para{Economics}
The \bsky network is currently fueled by the \bskyCo, enthusiasts, and early adopters.
Our analysis of account and Feed Generator descriptions suggests that multiple users raise money via dedicated services (\eg \url{patreon.com}, \url{ko-fi.com}) or point to alternative services hosting the users' and compensating creators (\eg \url{youtube.com}, \url{twitch.tv}).
\bskyCo does not currently suggest introducing advertisements.
However, our analysis of multiple forums and discussions around \bsky suggests that multiple creators consider introducing ads as posts included in their feed or Feed Generators. 
In the long run, the ecosystem will require economic incentives to become sustainable and compete with centralized platforms such as Twitter/X that recently started sharing its revenue with content creators~\cite{trevenue}.

\begin{acks}
	The authors would like to thank the anonymous referees for their valuable comments and helpful suggestions.
	This work was supported by the \grantsponsor{DFG}{German Research Foundation (DFG)}{https://www.dfg.de/}
	within the Collaborative Research Center (CRC) \grantnum[https://gepris.dfg.de/gepris/projekt/210487104]{DFG}{SFB 1053: MAKI}.
\end{acks}

\bibliographystyle{ACM-Reference-Format}
\balance
\bibliography{references}


\begin{thebibliography}{37}


\ifx \showCODEN    \undefined \def \showCODEN     #1{\unskip}     \fi
\ifx \showDOI      \undefined \def \showDOI       #1{#1}\fi
\ifx \showISBNx    \undefined \def \showISBNx     #1{\unskip}     \fi
\ifx \showISBNxiii \undefined \def \showISBNxiii  #1{\unskip}     \fi
\ifx \showISSN     \undefined \def \showISSN      #1{\unskip}     \fi
\ifx \showLCCN     \undefined \def \showLCCN      #1{\unskip}     \fi
\ifx \shownote     \undefined \def \shownote      #1{#1}          \fi
\ifx \showarticletitle \undefined \def \showarticletitle #1{#1}   \fi
\ifx \showURL      \undefined \def \showURL       {\relax}        \fi
\providecommand\bibfield[2]{#2}
\providecommand\bibinfo[2]{#2}
\providecommand\natexlab[1]{#1}
\providecommand\showeprint[2][]{arXiv:#2}

\bibitem[act({[n.\,d.]})]%
        {activitypub}
 \bibinfo{year}{[n.\,d.]}\natexlab{}.
\newblock \bibinfo{title}{{ActivityPub Specification}}.
\newblock \bibinfo{howpublished}{\url{https://www.w3.org/TR/activitypub/}}.
\newblock
\newblock
\shownote{Accessed on 15 May, 2024}.


\bibitem[tre({[n.\,d.]})]%
        {trevenue}
 \bibinfo{year}{[n.\,d.]}\natexlab{}.
\newblock \bibinfo{title}{{Ads Revenue Sharing}}.
\newblock
  \bibinfo{howpublished}{\url{https://help.twitter.com/en/using-x/creator-ads-revenue-sharing}}.
\newblock
\newblock
\shownote{Accessed on 15 May, 2024}.


\bibitem[mas({[n.\,d.]})]%
        {mastodon_users}
 \bibinfo{year}{[n.\,d.]}\natexlab{}.
\newblock \bibinfo{title}{{Mastodon Statistics}}.
\newblock \bibinfo{howpublished}{\url{https://mastodon-analytics.com/}}.
\newblock
\newblock
\shownote{Accessed on 15 May, 2024}.


\bibitem[nos({[n.\,d.]})]%
        {nostr}
 \bibinfo{year}{[n.\,d.]}\natexlab{}.
\newblock \bibinfo{title}{{NOSTR: A decentralized social network with a chance
  of working}}.
\newblock \bibinfo{howpublished}{\url{https://nostr.com/}}.
\newblock
\newblock
\shownote{Accessed on 29 April, 2024}.


\bibitem[bsk({[n.\,d.]})]%
        {bsky_blog_pt}
 \bibinfo{year}{[n.\,d.]}\natexlab{}.
\newblock \bibinfo{title}{{Perguntas Frequentes do Usuário Bluesky
  (Português)}}.
\newblock
  \bibinfo{howpublished}{\url{https://bsky.social/about/blog/04-10-2024-user-faq-br}}.
\newblock
\newblock
\shownote{Accessed on 15 May, 2024}.


\bibitem[twi({[n.\,d.]})]%
        {twitter_users}
 \bibinfo{year}{[n.\,d.]}\natexlab{}.
\newblock \bibinfo{title}{{X (Twitter) Statistics: How Many People Use X?}}
\newblock \bibinfo{howpublished}{\url{https://backlinko.com/twitter-users}}.
\newblock
\newblock
\shownote{Accessed on 15 May, 2024}.


\bibitem[Balduf et~al\mbox{.}(2023)]%
        {balduf2023cloud}
\bibfield{author}{\bibinfo{person}{Leonhard Balduf}, \bibinfo{person}{Maciej
  Korczy{\'n}ski}, \bibinfo{person}{Onur Ascigil}, \bibinfo{person}{Navin~V
  Keizer}, \bibinfo{person}{George Pavlou}, \bibinfo{person}{Bj{\"o}rn
  Scheuermann}, {and} \bibinfo{person}{Micha{\l} Kr{\'o}l}.}
  \bibinfo{year}{2023}\natexlab{}.
\newblock \showarticletitle{The Cloud Strikes Back: Investigating the
  Decentralization of IPFS}. In \bibinfo{booktitle}{\emph{Proceedings of the
  2023 ACM on Internet Measurement Conference}}. \bibinfo{pages}{391--405}.
\newblock


\bibitem[Bayer et~al\mbox{.}(2022)]%
        {bayer_eu_report}
\bibfield{author}{\bibinfo{person}{J Bayer}, \bibinfo{person}{Y Nosyk},
  \bibinfo{person}{O Hureau}, \bibinfo{person}{S Fernandez}, \bibinfo{person}{S
  Paulovics}, \bibinfo{person}{A Duda}, {and} \bibinfo{person}{M Korczynski}.}
  \bibinfo{year}{2022}\natexlab{}.
\newblock \bibinfo{booktitle}{\emph{Study on Domain Name System (DNS) abuse –
  Technical report. Appendix 1}}.
\newblock \bibinfo{publisher}{Publications Office of the European Union}.
\newblock
\urldef\tempurl%
\url{https://doi.org/doi/10.2759/473317}
\showDOI{\tempurl}


\bibitem[Bin~Zia et~al\mbox{.}(2024)]%
        {Binzia24}
\bibfield{author}{\bibinfo{person}{Haris Bin~Zia}, \bibinfo{person}{Jiahui He},
  \bibinfo{person}{Ignacio Castro}, {and} \bibinfo{person}{Gareth Tyson}.}
  \bibinfo{year}{2024}\natexlab{}.
\newblock \showarticletitle{Fediverse Migrations: A Study of User Account
  Portability on the Mastodon Social Network}. In
  \bibinfo{booktitle}{\emph{Proc. of ACM Internet Measurement Conference
  (IMC)}}.
\newblock


\bibitem[Bin~Zia et~al\mbox{.}(2022)]%
        {bin2022toxicity}
\bibfield{author}{\bibinfo{person}{Haris Bin~Zia}, \bibinfo{person}{Aravindh
  Raman}, \bibinfo{person}{Ignacio Castro}, \bibinfo{person}{Ishaku
  Hassan~Anaobi}, \bibinfo{person}{Emiliano De~Cristofaro},
  \bibinfo{person}{Nishanth Sastry}, {and} \bibinfo{person}{Gareth Tyson}.}
  \bibinfo{year}{2022}\natexlab{}.
\newblock \showarticletitle{Toxicity in the decentralized web and the potential
  for model sharing}.
\newblock \bibinfo{journal}{\emph{Proceedings of the ACM on Measurement and
  Analysis of Computing Systems}} \bibinfo{volume}{6}, \bibinfo{number}{2}
  (\bibinfo{year}{2022}), \bibinfo{pages}{1--25}.
\newblock


\bibitem[Cava et~al\mbox{.}(2021)]%
        {cava2021understanding}
\bibfield{author}{\bibinfo{person}{Lucio~La Cava}, \bibinfo{person}{Sergio
  Greco}, {and} \bibinfo{person}{Andrea Tagarelli}.}
  \bibinfo{year}{2021}\natexlab{}.
\newblock \showarticletitle{Understanding the growth of the Fediverse through
  the lens of Mastodon}.
\newblock \bibinfo{journal}{\emph{Applied Network Science}}
  \bibinfo{volume}{6} (\bibinfo{year}{2021}), \bibinfo{pages}{1--35}.
\newblock


\bibitem[Cava et~al\mbox{.}(2022a)]%
        {cava2022information}
\bibfield{author}{\bibinfo{person}{Lucio~La Cava}, \bibinfo{person}{Sergio
  Greco}, {and} \bibinfo{person}{Andrea Tagarelli}.}
  \bibinfo{year}{2022}\natexlab{a}.
\newblock \showarticletitle{Information consumption and boundary spanning in
  decentralized online social networks: the case of mastodon users}.
\newblock \bibinfo{journal}{\emph{Online Social Networks and Media}}
  \bibinfo{volume}{30} (\bibinfo{year}{2022}), \bibinfo{pages}{100220}.
\newblock


\bibitem[Cava et~al\mbox{.}(2022b)]%
        {cava2022network}
\bibfield{author}{\bibinfo{person}{Lucio~La Cava}, \bibinfo{person}{Sergio
  Greco}, {and} \bibinfo{person}{Andrea Tagarelli}.}
  \bibinfo{year}{2022}\natexlab{b}.
\newblock \showarticletitle{Network analysis of the information
  consumption-production dichotomy in mastodon user behaviors}. In
  \bibinfo{booktitle}{\emph{Proceedings of the International AAAI Conference on
  Web and Social Media}}, Vol.~\bibinfo{volume}{16}.
  \bibinfo{pages}{1378--1382}.
\newblock


\bibitem[{CNBC}({[n.\,d.]})]%
        {cnbc}
\bibfield{author}{\bibinfo{person}{{CNBC}}.}
  \bibinfo{year}{[n.\,d.]}\natexlab{}.
\newblock \bibinfo{title}{{Social media platform Bluesky attracts millions in
  Brazil after judge bans Musk’s X }}.
\newblock
  \bibinfo{howpublished}{\url{https://www.cnbc.com/2024/09/04/social-media-platform-bluesky-attracts-millions-in-brazil-after-judge-bans-musks-x-.html}}.
\newblock
\newblock
\shownote{Accessed on 11 Sep, 2024}.


\bibitem[Danilak et~al\mbox{.}(2021)]%
        {langdetect2021}
\bibfield{author}{\bibinfo{person}{Michal Danilak} {et~al\mbox{.}}}
  \bibinfo{year}{2021}\natexlab{}.
\newblock \bibinfo{title}{langdetect}.
\newblock \bibinfo{howpublished}{\url{https://pypi.org/project/langdetect/}}.
\newblock


\bibitem[Frier et~al\mbox{.}(2021)]%
        {frier2021why}
\bibfield{author}{\bibinfo{person}{Sarah Frier}, \bibinfo{person}{Naomi Nix},
  {and} \bibinfo{person}{Sarah Kopit}.} \bibinfo{year}{2021}\natexlab{}.
\newblock \showarticletitle{Why Free Speech on the Internet Isn’t Free for
  All.}
\newblock \bibinfo{journal}{\emph{International New York Times}}
  (\bibinfo{year}{2021}), \bibinfo{pages}{NA--NA}.
\newblock


\bibitem[Gribneau et~al\mbox{.}(2023)]%
        {did:web}
\bibfield{author}{\bibinfo{person}{Christian Gribneau},
  \bibinfo{person}{Michael Prorock}, \bibinfo{person}{Orie Steele},
  \bibinfo{person}{Oliver Terbu}, \bibinfo{person}{Mike Xu}, {and}
  \bibinfo{person}{Dmitri Zagidulin}.} \bibinfo{year}{2023}\natexlab{}.
\newblock \bibinfo{title}{DID WEB Method (did:web)}.
\newblock \bibinfo{howpublished}{\url{https://perma.cc/WB8M-8ECW}}.
\newblock


\bibitem[Guidi et~al\mbox{.}(2020)]%
        {guidi2020socioeconomic}
\bibfield{author}{\bibinfo{person}{Barbara Guidi}, \bibinfo{person}{Andrea
  Michienzi}, {and} \bibinfo{person}{Laura Ricci}.}
  \bibinfo{year}{2020}\natexlab{}.
\newblock \showarticletitle{A graph-based socioeconomic analysis of steemit}.
\newblock \bibinfo{journal}{\emph{IEEE Transactions on Computational Social
  Systems}} \bibinfo{volume}{8}, \bibinfo{number}{2} (\bibinfo{year}{2020}),
  \bibinfo{pages}{365--376}.
\newblock


\bibitem[Hassan et~al\mbox{.}(2021)]%
        {hassan2021exploring}
\bibfield{author}{\bibinfo{person}{Anaobi~Ishaku Hassan},
  \bibinfo{person}{Aravindh Raman}, \bibinfo{person}{Ignacio Castro},
  \bibinfo{person}{Haris~Bin Zia}, \bibinfo{person}{Emiliano De~Cristofaro},
  \bibinfo{person}{Nishanth Sastry}, {and} \bibinfo{person}{Gareth Tyson}.}
  \bibinfo{year}{2021}\natexlab{}.
\newblock \showarticletitle{Exploring content moderation in the decentralised
  web: The pleroma case}. In \bibinfo{booktitle}{\emph{Proceedings of the 17th
  International Conference on emerging Networking EXperiments and
  Technologies}}. \bibinfo{pages}{328--335}.
\newblock


\bibitem[He et~al\mbox{.}(2023)]%
        {he2023flocking}
\bibfield{author}{\bibinfo{person}{Jiahui He}, \bibinfo{person}{Haris~Bin Zia},
  \bibinfo{person}{Ignacio Castro}, \bibinfo{person}{Aravindh Raman},
  \bibinfo{person}{Nishanth Sastry}, {and} \bibinfo{person}{Gareth Tyson}.}
  \bibinfo{year}{2023}\natexlab{}.
\newblock \showarticletitle{Flocking to mastodon: Tracking the great twitter
  migration}. In \bibinfo{booktitle}{\emph{Proceedings of the 2023 ACM on
  Internet Measurement Conference}}. \bibinfo{pages}{111--123}.
\newblock


\bibitem[Holmgren et~al\mbox{.}(2023)]%
        {did:plc}
\bibfield{author}{\bibinfo{person}{Daniel Holmgren}, \bibinfo{person}{Bryan
  Newbold}, \bibinfo{person}{Devin Ivy}, {and} \bibinfo{person}{Jake Gold}.}
  \bibinfo{year}{2023}\natexlab{}.
\newblock \bibinfo{title}{{DID PLC Method (did:plc)}}.
\newblock
  \bibinfo{howpublished}{\url{https://github.com/did-method-plc/did-method-plc}}.
\newblock


\bibitem[Jaatun et~al\mbox{.}(2022)]%
        {jaatun2022}
\bibfield{author}{\bibinfo{person}{Lars~Andreassen Jaatun},
  \bibinfo{person}{Anders Ringen}, {and} \bibinfo{person}{Martin~Gilje
  Jaatun}.} \bibinfo{year}{2022}\natexlab{}.
\newblock \showarticletitle{Yet Another Blockchain-based Privacy-friendly
  Social Network}. In \bibinfo{booktitle}{\emph{2022 IEEE International
  Conference on Cloud Computing Technology and Science (CloudCom)}}. IEEE,
  \bibinfo{pages}{222--229}.
\newblock


\bibitem[Jeong et~al\mbox{.}(2024a)]%
        {jeong2024bluetempnettemporalmultinetworkdataset}
\bibfield{author}{\bibinfo{person}{Ujun Jeong}, \bibinfo{person}{Bohan Jiang},
  \bibinfo{person}{Zhen Tan}, \bibinfo{person}{H.~Russell Bernard}, {and}
  \bibinfo{person}{Huan Liu}.} \bibinfo{year}{2024}\natexlab{a}.
\newblock \bibinfo{title}{BlueTempNet: A Temporal Multi-network Dataset of
  Social Interactions in Bluesky Social}.
\newblock
\newblock
\showeprint[arxiv]{2407.17451}~[cs.SI]
\urldef\tempurl%
\url{https://arxiv.org/abs/2407.17451}
\showURL{%
\tempurl}


\bibitem[Jeong et~al\mbox{.}(2024b)]%
        {jeong2024usermigrationmultiplesocial}
\bibfield{author}{\bibinfo{person}{Ujun Jeong}, \bibinfo{person}{Ayushi
  Nirmal}, \bibinfo{person}{Kritshekhar Jha}, \bibinfo{person}{Susan~Xu Tang},
  \bibinfo{person}{H.~Russell Bernard}, {and} \bibinfo{person}{Huan Liu}.}
  \bibinfo{year}{2024}\natexlab{b}.
\newblock \bibinfo{title}{User Migration across Multiple Social Media
  Platforms}.
\newblock
\newblock
\showeprint[arxiv]{2309.12613}~[cs.SI]
\urldef\tempurl%
\url{https://arxiv.org/abs/2309.12613}
\showURL{%
\tempurl}


\bibitem[Kleppmann et~al\mbox{.}(2024)]%
        {kleppmann2024bluesky}
\bibfield{author}{\bibinfo{person}{Martin Kleppmann}, \bibinfo{person}{Paul
  Frazee}, \bibinfo{person}{Jake Gold}, \bibinfo{person}{Jay Graber},
  \bibinfo{person}{Daniel Holmgren}, \bibinfo{person}{Devin Ivy},
  \bibinfo{person}{Jeromy Johnson}, \bibinfo{person}{Bryan Newbold}, {and}
  \bibinfo{person}{Jaz Volpert}.} \bibinfo{year}{2024}\natexlab{}.
\newblock \showarticletitle{Bluesky and the AT Protocol: Usable Decentralized
  Social Media}.
\newblock \bibinfo{journal}{\emph{arXiv preprint arXiv:2402.03239}}
  (\bibinfo{year}{2024}).
\newblock


\bibitem[Le~Pochat et~al\mbox{.}(2019)]%
        {pochat2018tranco}
\bibfield{author}{\bibinfo{person}{Victor Le~Pochat}, \bibinfo{person}{Tom
  Van~Goethem}, \bibinfo{person}{Samaneh Tajalizadehkhoob},
  \bibinfo{person}{Maciej Korczy\'nski}, {and} \bibinfo{person}{Wouter
  Joosen}.} \bibinfo{year}{2019}\natexlab{}.
\newblock \showarticletitle{Tranco: A Research-Oriented Top Sites Ranking
  Hardened Against Manipulation}. In \bibinfo{booktitle}{\emph{NDSS}}.
\newblock


\bibitem[McQuistin et~al\mbox{.}(2023)]%
        {mcquistin2023first}
\bibfield{author}{\bibinfo{person}{Stephen McQuistin}, \bibinfo{person}{Peter
  Snyder}, \bibinfo{person}{Colin Perkins}, \bibinfo{person}{Hamed Haddadi},
  {and} \bibinfo{person}{Gareth Tyson}.} \bibinfo{year}{2023}\natexlab{}.
\newblock \showarticletitle{A first look at the privacy harms of the public
  suffix list}. In \bibinfo{booktitle}{\emph{Proceedings of the 2023 ACM on
  Internet Measurement Conference}}. \bibinfo{pages}{383--390}.
\newblock


\bibitem[Quelle and Bovet(2024)]%
        {quelle2024blueskynetworktopologypolarization}
\bibfield{author}{\bibinfo{person}{Dorian Quelle} {and}
  \bibinfo{person}{Alexandre Bovet}.} \bibinfo{year}{2024}\natexlab{}.
\newblock \bibinfo{title}{Bluesky: Network Topology, Polarization, and
  Algorithmic Curation}.
\newblock
\newblock
\showeprint[arxiv]{2405.17571}~[cs.SI]
\urldef\tempurl%
\url{https://arxiv.org/abs/2405.17571}
\showURL{%
\tempurl}


\bibitem[Raman et~al\mbox{.}(2019)]%
        {raman2019challenges}
\bibfield{author}{\bibinfo{person}{Aravindh Raman}, \bibinfo{person}{Sagar
  Joglekar}, \bibinfo{person}{Emiliano~De Cristofaro},
  \bibinfo{person}{Nishanth Sastry}, {and} \bibinfo{person}{Gareth Tyson}.}
  \bibinfo{year}{2019}\natexlab{}.
\newblock \showarticletitle{Challenges in the decentralised web: The mastodon
  case}. In \bibinfo{booktitle}{\emph{Proceedings of the internet measurement
  conference}}. \bibinfo{pages}{217--229}.
\newblock


\bibitem[Rozenshtein(2023)]%
        {rozenshtein2023moderating}
\bibfield{author}{\bibinfo{person}{Alan~Z Rozenshtein}.}
  \bibinfo{year}{2023}\natexlab{}.
\newblock \showarticletitle{Moderating the fediverse: Content moderation on
  distributed social media}.
\newblock \bibinfo{journal}{\emph{J. Free Speech L.}}  \bibinfo{volume}{3}
  (\bibinfo{year}{2023}), \bibinfo{pages}{217}.
\newblock


\bibitem[Satariano(2021)]%
        {satariano2021facebook}
\bibfield{author}{\bibinfo{person}{Adam Satariano}.}
  \bibinfo{year}{2021}\natexlab{}.
\newblock \showarticletitle{Facebook Hearing Strengthens Calls for Regulation
  in Europe.}
\newblock \bibinfo{journal}{\emph{International New York Times}}
  (\bibinfo{year}{2021}), \bibinfo{pages}{NA--NA}.
\newblock


\bibitem[Sporny et~al\mbox{.}(2022)]%
        {did}
\bibfield{author}{\bibinfo{person}{Manu Sporny}, \bibinfo{person}{Dave
  Longley}, \bibinfo{person}{Markus Sabadello}, \bibinfo{person}{Drummond
  Reed}, \bibinfo{person}{Orie Steele}, {and} \bibinfo{person}{Christopher
  Allen}.} \bibinfo{year}{2022}\natexlab{}.
\newblock \bibinfo{title}{{Decentralized Identifiers (DIDs) v1.0}}.
\newblock \bibinfo{howpublished}{\url{https://www.w3.org/TR/did-core/}}.
\newblock


\bibitem[Tarr et~al\mbox{.}(2019)]%
        {SSB}
\bibfield{author}{\bibinfo{person}{Dominic Tarr}, \bibinfo{person}{Erick
  Lavoie}, \bibinfo{person}{Aljoscha Meyer}, {and} \bibinfo{person}{Christian
  Tschudin}.} \bibinfo{year}{2019}\natexlab{}.
\newblock \showarticletitle{Secure Scuttlebutt: An Identity-Centric Protocol
  for Subjective and Decentralized Applications} \emph{(\bibinfo{series}{ICN
  '19})}. \bibinfo{publisher}{Association for Computing Machinery},
  \bibinfo{address}{New York, NY, USA}, \bibinfo{pages}{1–11}.
\newblock
\showISBNx{9781450369701}
\urldef\tempurl%
\url{https://doi.org/10.1145/3357150.3357396}
\showDOI{\tempurl}


\bibitem[Townsend and Wallace(2017)]%
        {townsend2017ethics}
\bibfield{author}{\bibinfo{person}{Leanne Townsend} {and}
  \bibinfo{person}{Claire Wallace}.} \bibinfo{year}{2017}\natexlab{}.
\newblock \showarticletitle{The ethics of using social media data in research:
  A new framework}.
\newblock In \bibinfo{booktitle}{\emph{The ethics of online research}}.
  \bibinfo{publisher}{Emerald Publishing Limited}, \bibinfo{pages}{189--207}.
\newblock


\bibitem[Wei et~al\mbox{.}(2024)]%
        {wei2024eternal}
\bibfield{author}{\bibinfo{person}{Yiluo Wei}, \bibinfo{person}{Dennis
  Trautwein}, \bibinfo{person}{Yiannis Psaras}, \bibinfo{person}{Ignacio
  Castro}, \bibinfo{person}{Will Scott}, \bibinfo{person}{Aravindh Raman},
  {and} \bibinfo{person}{Gareth Tyson}.} \bibinfo{year}{2024}\natexlab{}.
\newblock \showarticletitle{The Eternal Tussle: Exploring the Role of
  Centralization in $\{$IPFS$\}$}. In \bibinfo{booktitle}{\emph{21st USENIX
  Symposium on Networked Systems Design and Implementation (NSDI 24)}}.
  \bibinfo{pages}{441--454}.
\newblock


\bibitem[Whitten and Tygar(1999)]%
        {whitten1999johnny}
\bibfield{author}{\bibinfo{person}{Alma Whitten} {and} \bibinfo{person}{J~Doug
  Tygar}.} \bibinfo{year}{1999}\natexlab{}.
\newblock \showarticletitle{Why Johnny Can't Encrypt: A Usability Evaluation of
  PGP 5.0.}. In \bibinfo{booktitle}{\emph{USENIX security symposium}},
  Vol.~\bibinfo{volume}{348}. \bibinfo{pages}{169--184}.
\newblock


\bibitem[Zuo et~al\mbox{.}(2023)]%
        {memo}
\bibfield{author}{\bibinfo{person}{Wenrui Zuo}, \bibinfo{person}{Aravindh
  Raman}, \bibinfo{person}{Raul~J Mondrag\'{o}n}, {and} \bibinfo{person}{Gareth
  Tyson}.} \bibinfo{year}{2023}\natexlab{}.
\newblock \showarticletitle{Set in Stone: Analysis of an Immutable Web3 Social
  Media Platform} \emph{(\bibinfo{series}{WWW '23})}.
  \bibinfo{publisher}{Association for Computing Machinery},
  \bibinfo{address}{New York, NY, USA}, \bibinfo{pages}{1865–1874}.
\newblock
\showISBNx{9781450394161}
\urldef\tempurl%
\url{https://doi.org/10.1145/3543507.3583510}
\showDOI{\tempurl}


\end{thebibliography}

\appendix

\section{Additional Data}\label{app:data}
We provide additional information on Feed Generator As a Service platforms (\Cref{tab:feed_comparison}) and the complete table of reaction times of labelers to posts published via the Firehose (\Cref{tab:labelers_firehose_post_label_time_diff_full}).
\begin{table*}[ht]
	\centering
	\caption{Comparing the top 5 feed generator builders/services.}
	\label{tab:feed_comparison}
	{\small%
		\begin{tabular}{@{}llllll@{}}
\toprule
\textbf{Feature} & \textbf{Skyfeed} & \textbf{Bluefeed} & \textbf{Blueskyfeeds} & \textbf{goodfeeds} & \textbf{Blueskyfeedcreator} \\ \midrule
\textbf{Inputs} & & & & & \\
Whole network & \checkmark & \checkmark &  & \checkmark &  \checkmark \\
Tags & \checkmark & \checkmark & \checkmark &  &\checkmark  \\
Single user & \checkmark & \checkmark & \checkmark & \checkmark  & \checkmark  \\
List & \checkmark &  & \checkmark & \checkmark  &  \checkmark \\
Feed & \checkmark & \checkmark &  &  &  \\
Single post & \checkmark & \checkmark & \checkmark &  &  \\
Labels & \checkmark & \checkmark &  &  &  \\
Token &  &  & \checkmark &  &  \\
Segment &  &  & \checkmark &  &  \\ \midrule

\textbf{Filters} \\
Item & \checkmark &  &  &  & \checkmark \\
Labels & \checkmark & \checkmark & \checkmark  &   & \checkmark  \\
Image count &\checkmark  &  &  &  &  \\
Link count &\checkmark  &  &  &  &  \\
Repost count &\checkmark  &  &  &  &  \\
Embed &\checkmark  &  &  &  &  \\
Duplicate &\checkmark  &  &  &  &  \\
List of users &\checkmark  &  &  \checkmark&  &\checkmark  \\
Language &\checkmark  &  &\checkmark  &  & \checkmark  \\ \midrule
\textbf{Regex} \\
Text & \checkmark &  &  &  &  \\
Image Alt & \checkmark &  &  &  &  \\
Link & \checkmark &  &  &  &  \\ \midrule
\textbf{Other Features} \\
Number of Feeds & 35,415 &  2,302 & 1,797 & 929 & 158 \\
Paid or Free & free & free & free & free & free \& paid \\ \bottomrule
\end{tabular}
\unskip%
	}
\end{table*}

\begin{table*}[ht]
	\centering
	\caption{Reaction time of Labelers to Posts Published via the Firehose.}
	\label{tab:labelers_firehose_post_label_time_diff_full}
	{\small%
\begin{tabular}{rlp{4cm}rrrrr}
  \toprule
  && \multicolumn{4}{c}{Labels Applied} & \multicolumn{2}{c}{Reaction Time [s]} \\ 
\cmidrule(lr){3-6}\cmidrule(lr){7-8}Rank & DID & Top Values & \# Unique & \# Total & Share (\%) & Median & IQD \\ 
 \midrule
  1 & did:plc:wp7hxfjl5l4zlptn7y6774lk & no-alt-text, non-alt-text, mis-alt-text & $4$ & $1{,}360{,}224$ & $72.91$ & $0.58$ & $0.10$ \\ 
    2 & did:plc:ar7c4by46qjdydhdevvrndac & porn, sexual, nudity & $32$ & $279{,}002$ & $14.95$ & $1.76$ & $0.70$ \\ 
    3 & did:plc:newitj5jo3uel7o4mnf3vj2o & twitter-screenshot, bluesky-screenshot, uncategorised-screenshot & $14$ & $76{,}599$ & $4.11$ & $3.70$ & $3.81$ \\ 
    4 & did:plc:mjyeurqmqjeexbgigk3yytvb & tenor-gif, tenor-gif-no-text & $2$ & $73{,}875$ & $3.96$ & $0.35$ & $0.20$ \\ 
    5 & did:plc:bpkpvmwpd3nr2ry4btt55ack & ai-imagery & $1$ & $56{,}517$ & $3.03$ & $0.82$ & $0.21$ \\ 
    6 & did:plc:fcikraffwejtuqffifeykcml & shadowbringers, endwalker, dawntrail & $6$ & $10{,}024$ & $0.54$ & $2.07$ & $0.82$ \\ 
    7 & did:plc:3eivfiql4memqxkryeu4tqnk & ai-related-content, spoiler, test-label & $3$ & $7{,}646$ & $0.41$ & $1.32$ & $0.78$ \\ 
    8 & did:plc:j67mwmangcbxch7knfm7jo2b & trolling, transphobia, racial-intolerance & $13$ & $876$ & $0.05$ & $13{,}911.90$ & $53{,}085.19$ \\ 
    9 & did:plc:vrjubqujt3v46z5poehh4qfg & pup, fatfur, diaper & $18$ & $631$ & $0.03$ & $34{,}408.43$ & $65{,}282.36$ \\ 
   10 & did:plc:3ehw5dwwptcy3xuzugwq2u6t & beans & $1$ & $49$ & $< 0.01$ & $90.39$ & $5{,}089.05$ \\ 
   11 & did:plc:skibpmllbhxvbvwgtjxl3uao & simping, bad-selfies, cringe & $5$ & $32$ & $< 0.01$ & $70{,}413.53$ & $121{,}503.24$ \\ 
   12 & did:plc:olmiw2wkm3qoxinal7w5fbnl & lowquality, shorturl, unknownsource & $6$ & $26$ & $< 0.01$ & $104{,}584.57$ & $236{,}752.45$ \\ 
   13 & did:plc:e4elbtctnfqocyfcml6h2lf7 & alf, sensual-alf, the-format & $3$ & $18$ & $< 0.01$ & $38{,}417.71$ & $61{,}154.18$ \\ 
   14 & did:plc:exlb5xx2t4pgtjqzdm6ntsgh & severity-alert-blurs-content, severity-alert-blurs-media, severity-alert-blurs-none & $9$ & $18$ & $< 0.01$ & $937.55$ & $584.76$ \\ 
   15 & did:plc:4vf7tgwlg6edds2g2nixyjda & spam-aff-ja, spam, porn & $4$ & $16$ & $< 0.01$ & $534{,}935.10$ & $429{,}626.79$ \\ 
   16 & did:plc:gcbmhqcuvuoz7jgmlanabiuv & so-true, epic, based & $4$ & $16$ & $< 0.01$ & $526.03$ & $3{,}413.47$ \\ 
   17 & did:plc:5o2g6wwchb3tgwrhl2atauzu & !warn, threat, triggerwarning & $10$ & $14$ & $< 0.01$ & $109{,}931.10$ & $373{,}967.40$ \\ 
   18 & did:plc:36inn6r2ttwfrt6tpywsjcmt & coulro, arachno, lepidoptero & $6$ & $11$ & $< 0.01$ & $260{,}511.95$ & $297{,}492.05$ \\ 
   19 & did:plc:cnn3jrtucivembf66xe6fdfs & neutral-pro-discourse, anti-discourse & $2$ & $10$ & $< 0.01$ & $2{,}120.64$ & $47{,}340.94$ \\ 
   20 & did:plc:mcskx665cnmnkgqnunk6lkrk & spoilers, !no-promote, !no-unauthenticated & $3$ & $4$ & $< 0.01$ & $1{,}585{,}404.55$ & $3{,}100{,}279.22$ \\ 
   21 & did:plc:z2i5ah5elywxdcr64i7xai3z & nipps, no-church, non-handshake & $3$ & $4$ & $< 0.01$ & $154{,}416.53$ & $95{,}557.08$ \\ 
   22 & did:plc:7fkqmr7dfu6vanyxvjtloos3 & !warn, porn, spam & $3$ & $3$ & $< 0.01$ & $5{,}203.95$ & $95{,}853.62$ \\ 
   23 & did:plc:j2zujaxuq33c7nbcqyvgvyvk & amplifying-disinfo & $1$ & $3$ & $< 0.01$ & $5{,}445.06$ & $9{,}348.35$ \\ 
   24 & did:plc:hxgctysbwhc3bap3a5c7gdu3 & beanhate, feature-scold & $2$ & $2$ & $< 0.01$ & $5{,}900.41$ & $4{,}489.93$ \\ 
   \bottomrule
\end{tabular}
\unskip%
	}
\end{table*}

\section{Ethics} \label{sec:ethics}
We believe that the benefits of our research significantly outweigh potential harms.
This work helps understand the implications of opening and decentralizing social network platforms.
We take multiple actions to minimize any potential harm.
We exclusively collect publicly available information and follow well established ethical procedures for social data~\cite{townsend2017ethics}.
We make no attempt to link activities to other accounts or real-world identifies and the collected data is stored securely within a university silo, and no external access is given.
Prior to initiating our scans, we contacted the Bluesky team to agree upon a scanning rate that would not disrupt the normal functioning of their service.
Additionally, and to ensure a minimal impact, we implement a solution that exclusively downloads repositories if their content has changed since the last snapshot. 
We identified issues preventing this originally and upstreamed fixes to the Bluesky open-source projects.

\leo{TODO: We also keep repos of deleted accounts/accounts taken down. Is this ethically safe? In particular, do they not have a right to be forgotten or something?}
\michal{I think it's ok. It'd be a problem if we kept them while someone asked us to delete it but let's confirm.}
\ignacio{may be we delete them once we find them to be deleted but we collect the stats?}

When analyzing the public endpoints of community labeling services we deduce their IP addresses while subscribing to them for labels, which is intended behavior.
Analyses of the IP addresses themselves happened locally on university machines.

\end{document}